\documentclass[11pt]{article}

\usepackage{amsmath,amsfonts,amssymb}
\usepackage{bbold}

\numberwithin{equation}{section}
\newcommand{\ns}{\normalsize}
\newcommand{\unitop}{\mathbb{1}}

\newcommand{\ax}{\alpha}
\newcommand{\bx}{\beta}

\newcommand{\dx}{\delta}

\newcommand{\lx}{\lambda}
\newcommand{\gx}{\gamma}

\newcommand{\Gx}{\Gamma}

\newcommand{\e}{\epsilon}
\newcommand{\px}{\varphi}
\newcommand{\kx}{\kappa}

\newcommand{\cK}{\mathcal{K}}

\newcommand{\W}{\mathcal W}
\newcommand{\V}{\mathcal V}

\newcommand{\wg}{\wedge}
\newcommand{\rep}[1]{\mathbf{#1}}

\newcommand{\nn}{\nonumber}

\newcommand{\DET}{\mathrm{det} \,}
\newcommand{\g}{g}   

\newcommand{\bgd}[1]{\mathring{#1}}

\begin{document}


\begin{titlepage}

\title{
   \hfill{\ns hep-th/0412006\\}
   \vskip 2cm
   {\Large\bf M-theory Compactifications on Manifolds with $G_2$
   Structure}\\[0.5cm]} 
\author{
{\ns\large
    Thomas House\footnote{email: thomash@sussex.ac.uk}}~,
  {\ns and Andrei Micu\footnote{email: A.Micu@sussex.ac.uk; on leave
      from IFIN-HH Bucharest.}} 
  \\[0.5cm]
   {\it\ns Department of Physics and Astronomy, University of Sussex}\\
   {\ns Brighton BN1 9QH, UK}}
\date{}

\maketitle

\begin{abstract}\noindent
  
  In this paper we study M-theory compactifications on manifolds of
  $G_2$ structure. By computing the gravitino mass term in four
  dimensions we derive the general form for the superpotential which
  appears in such compactifications and show that beside the normal
  flux term there is a term which appears only for non-minimal $G_2$
  structure. We further apply these results to compactifications on
  manifolds with weak $G_2$ holonomy and make a couple of statements
  regarding the deformation space of such manifolds. Finally we show
  that the superpotential derived from fermionic terms leads to the
  potential that can be derived from the explicit compactification,
  thus strengthening the conjectures we make about the space of
  deformations of manifolds with weak $G_2$ holonomy.

\end{abstract}

\thispagestyle{empty}

\end{titlepage}


\section{Introduction}

Low-energy M-theory solutions with seven compact and four large
dimensions that yield $N=1$ supergravities play an important role in
the `landscape' of string and M-theory vacua.  As is always the case,
lower dimensional supersymmetry imposes strong constraints on the
internal manifold. In particular this should posses at least one
globally defined spinor or---equivalently---have its structure group
contained in $G_2 \subset SO(7)$ \cite{DNP}. If one also requires that
the four-dimensional space-time be flat then the internal manifold
must be Ricci flat and have holonomy group contained in $G_2$
\cite{PT}.

If, on the other hand, one allows for a curved four-dimensional
space-time whilst still preserving maximal symmetry and
supersymmetry in four dimensions, then the external space becomes
AdS. The internal manifold then need not be Ricci flat and will
typically be an Einstein space of constant positive curvature.
Solutions of this type were extensively considered after the discovery
of $D=11$ supergravity and are known as Freund--Rubin spontaneous
compactifications \cite{FR}.  Much of this earlier work is reviewed in
\cite{DNP}.  Using the elegant notions of $G$-structures, the internal
spaces for these solutions that support a single globally defined
spinor will be of the so-called weak $G_2$ type \cite{BJ, ADHL}.

The compactifications described above suffer from two major problems,
which are common to most string/M-theory compactifications. First of
all, they give rise to many moduli in four dimensions. These are
massless scalar fields which are exact flat directions of the
potential and parameterise a huge vacuum degeneracy. Second, the
spectrum of fermions in four dimensions is in general non-chiral --
and thus not suitable for particle physics phenomenolygy -- due
to a quite general index theorem for smooth seven-dimensional spaces
\cite{W1}. 

As became customary in the last period, the first problem can be
tackled by turning on non-trivial background fluxes which induce a
(super)potential for the scalar fields \cite{BW,Acharya1}. There was
also recent work on moduli fixing for M-theory on $G_2$ manifolds
including non-perturbative contributions in \cite{BAS}.

The resolution of the second problem came with the advent of dualities
and the realisation that chiral fermions appear at singular conical
points in the seven-dimensional manifold \cite{AW,AcW}, together with
a mechanism for the cancellation of the appropriate anomalies
\cite{W2}.\footnote{Such compactifications turn out to be dual to
  intersecting brane models in the context of type IIA string theory,
  as originally discussed in \cite{CSU1, CSU2}.} Conical singularities
are quite well understood in the non-compact case \cite{ADHL}, but so
far it has proved difficult to construct compact $G_2$ holonomy
manifolds with point-like singularities. This is partly because the
standard way in which such manifolds are constructed is by the
orbifolding of seven-tori followed by the blowing up of the orbifold
singularities \cite{Joyce,LM,BL}, which does not naturally produce
point-like singularities.

The known examples of compact manifolds with codimension seven
singularities are not in fact manifolds with $G_2$ holonomy, but weak
$G_2$ manifolds, and it is known \cite{ADHL} that chiral fermions can
also emerge from singular weak $G_2$ manifolds. These manifolds can be
constructed by adding an extra compact dimension to weak $SU(3)$
spaces to form a `lemon' with two conical singularities \cite{BM}.
Although having two singularities presents phenomenological problems,
spaces such as these may play a role as simple models, and motivate
further study into general features of weak $G_2$ manifolds.

The presence of background fluxes---when not perturbatively small---in
M-theory as well as in string compactifications induces a back-reaction
on the underlying geometry so that manifolds of $G_2$ holonomy are no
longer solutions to the equations of motion. In turn the resulting
geometry can be described in terms of the more general concept of
$G$-structures \cite{BJ}\cite{BDS}--\cite{GMSW}.
Compactifications on manifolds with non-trivial $G$ structure are
poorly understood, however, mainly because of the lack of knowledge of
the deformation space of these manifolds and it will be our purpose in
this paper to clarify some aspects of these compactifications for the
particular case of manifolds with $G_2$ structure. In several works it
was proposed that a superpotential which depends on the structure is
generated, but in most cases the expression of these superpotentials
in terms of the low energy fields and their relevance for moduli
stabilisation was not possible to compute due to the lack of
understanding of the low energy degrees of freedom which appear in
such compactification.

A different route to understand $G$-structures in string and M-theory
compactifications is via dualities \cite{GLMW}--\cite{AM}. In this way
it is possible to deduce certain properties of the moduli space of
these manifolds. To our knowledge this was only done in \cite{GLMW}
where the low energy action was also computed. For such models an
explicit expression for the superpotential in terms of the low energy
fields can be found \cite{GLM} and thus it is possible to say
something about the dynamics (and in particular moduli stabilisation).
The main assumption in these papers was that the fluxes/intrinsic
torsion are perturbatively small and that in such a regime the moduli
space is similar to the moduli space of some Calabi--Yau manifold. The
generalisation of this conjecture for large fluxes/intrinsic torsion
is, however, not known.

In a recent work, another conjecture about the deformation space of
nearly-K\"ahler manifolds (known also as manifolds of weak $SU(3)$
holonomy) was made \cite{AM}. In this paper it was proposed that the
weak $SU(3)$ conditions impose strong constraints on the space of
possible perturbations of such manifolds and only size deformations of
these manifolds are allowed.

The purpose of this paper is twofold. In the first instance we study
some general aspects of M-theory compactifications on manifolds with
$G_2$ structure. By compactifying certain fermionic terms we derive
the general form of the superpotential which appears in such
compactifications. The formula we obtain generalises in a natural way
the result obtained in \cite{BW} for just $F_4$ fluxes based on the
general analysis in \cite{GVW,SG}. Motivated by
the fact that the manifolds with weak $SU(3)$ holonomy are tightly
constrained, in the second part of this paper we study their
seven-dimensional relatives, namely manifolds with weak $G_2$
holonomy. Such manifolds have the property that the intrinsic
(con)torsion is a singlet under the structure group $G_2$ and it turns
out that one can infer enough about their internal structure to allow
us to derive the low-energy effective action for M-theory compactified
on such manifolds. In the end we link the two parts of the paper by
showing that the general formula for the superpotential derived in
general for any $G_2$ structure produces the correct result for the
manifolds with weak $G_2$ holonomy in that the potential which is
obtained from the compactification can be also obtained from this
superpotential when inserted in the standard $N=1$ formula.

M-theory on manifolds with restricted structure group was also studied
in \cite{BJ,MS}. Here it was shown among other things that if the
structure group is exactly $G_2$ the only supersymmetric solution is
given by the Freund--Rubin compactification \cite{DNP} and that the
internal manifold must be of weak $G_2$ type. In this paper we go one
step further and present the generic form of the low-energy theory
obtained in such compactifications. It is important to observe that
the four-dimensional ground state in this case is AdS which together
with the presence of a non-trivial flux along the four space-time
dimensions changes the definition of the mass operators for the fields
which appear in the low-energy theory. This should tell us that the
appropriate AdS massless modes no longer appear when the fields are
expanded in harmonic forms, but in forms which satisfy
\begin{equation}
  d \ax_{(3)} = - \tau * \ax_{(3)} \; ,
\end{equation}
where $\tau$ measures the intrinsic torsion of the manifold with weak
$G_2$ holonomy. It will also turn out that these are precisely the
variations of the $G_2$ structure which are compatible with the weak
$G_2$ conditions. Thus one can see in the same way as on manifolds
with $G_2$ holonomy that the modes coming from the matter fields
combine with the modes coming from the metric into the complex scalars
$a_3 + i \px$.

The organisation of the paper is the following. In section 2 we
present our conventions for the M-theory action and $G_2$ structures.
In section 3 we derive the gravitino mass term which appears in
compactifications on manifolds with $G_2$ structure which will give us
the K\"ahler potential and the superpotential for the chiral fields
coupled to supergravity in four dimensions. In section 4 we discuss
weak $G_2$ manifolds and concentrate on their deformation space. Using
this information we perform in section 5 the Kaluza--Klein reduction
on manifolds with weak $G_2$ holonomy and show agreement with the
general results derived in section 3 for the superpotential and
K\"ahler potential. In section 6 we present our conclusions. For
completeness we have also assembled a couple of appendices where we
present our conventions and the most useful relations on manifolds
with $G_2$ structure which we use throughout the paper. Here we have also
included some of the technical details which were skipped in the main
part of the paper.

\section{Preliminaries}

\subsection{Action and Ansatz}

We start by introducing our M-theory conventions which we will use
throughout the paper. As it is well-known, the low-energy
approximation of M-theory is given by the eleven-dimensional
supergravity which describes the dynamics of the $N=1$ supermultiplet
in eleven dimensions. This contains the metric $\hat g_{MN}$ and an
antisymmetric tensor field $\hat A_{MNP}$ as bosonic components and the
gravitino $\hat \Psi_M$, which is a Majorana spinor in eleven
dimensions, as their fermionic superpartner. Hats are 
used to distinguish the eleven-dimensional fields from the
four- and seven-dimensional ones that we will introduce later on. 
We use $D$ for the spinor covariant derivative. The indices
$M, N \ldots = 0,1, \ldots ,10$ denote curved eleven-dimensional indices.
The supergravity action can then be written as in \cite{BBS}
\begin{eqnarray}
  \label{S11}
  S & = & \frac{1}{2} \int_{M_{11}} \sqrt{-{\hat g}} \; d^{11} x 
  \bigg[ \hat R -  \overline{\hat \Psi}_M {\hat \Gamma}^{MNP} D_N
  \hat \Psi_P - \frac12 \hat F_4 \wedge * \hat F^4 \bigg] \nn \\[2mm]
  & & - \frac{1}{192} \int_{M_{11}} \sqrt{-{\hat g}} \; d^{11} x
  \overline{\hat \Psi}_M {\hat \Gamma}^{MNPQRS} \hat \Psi_N (\hat F_4)_{PQRS} 
  - \frac12 \int_{M_{11}} \hat F_4 \wedge * \hat C_4  \\[2mm]
  && - \frac{1}{12}  \int_{M_{11}} \hat F_4 \wedge \hat F_4
  \wedge \hat A_3 \; , \nn
\end{eqnarray}
where we have set the eleven-dimensional Newton's constant to unity
and denoted
\begin{equation}
  (\hat C_4)_{MNPQ} = 3 \overline{\hat \Psi}_{[M} {\hat \Gamma}_{NP} \hat
  \Psi_{Q]} \; .
\end{equation}
The spinor conjugation is defined 
to be $\overline{\hat \Psi}_M = \hat \Psi_M^\dagger \Gx^0 $.  We have
ignored the four-fermionic terms, which play no role in our analysis,
and kept only bilinear terms in the gravitino field $\hat \Psi_M$. The
action \eqref{S11} is invariant under the usual supersymmetry
transformations. For the gravitino this takes the form
\begin{equation}
  \label{susytrs}
  \dx \Psi_M = D_M \e + \frac{1}{288} 
  \left( {\hat \Gamma_M}^{\ \ NPQR} - 8 \delta_M^N \hat \Gamma^{NPQ}
  \right) (\hat F_4)_{NPQR} \; \e \; ,
\end{equation}
where $\e$ is the Majorana spinor parameterising the supersymmetry
transformation. The conventions for the Dirac matrices ${ \Gamma_M }$
are given in appendix \ref{gammaproperties}.

One of the aims of this paper is to compactify the action \eqref{S11}
to give a four-dimensional theory. We achieve this by splitting the
eleven-dimensional space as
\begin{equation}
  \label{split}
  M_{11} = M_4 \times K_7 \; .
\end{equation}
As a consequence the metric in eleven dimensions decomposes into a
direct product of a Minkowski signature metric on the four-dimensional
space $M_4$ and an Euclidean one on the internal manifold $K_7$
\begin{equation}
  \label{redAnsg}
  ds_{11}^2 =
  {\hat g}_{\mu \nu} dx^\mu dx^\nu + {\hat g}_{mn} dy^m dy^n \; ,
\end{equation}
where $\mu, \nu = 0, \ldots, 3$ are curved indices for the four-dimensional
Minkowski space and $m, n, \ldots = 4, \ldots 10$ are curved indices
over the seven-dimensional compact space.

In order to compute fermionic terms in the four-dimensional effective
action one needs to specify how the gamma matrices ${\hat\Gx}_M$ decompose
under \eqref{split}
\begin{equation}
  \begin{aligned}
    {\hat\Gx}_{\mu} & = {\hat\gx}_{\mu} \otimes \unitop \; ,\\
    {\hat\Gx}_m & = \gx \otimes {\hat\gx}_m \; .    
  \end{aligned}
\end{equation}
We deal with the conventions for $\{ \hat \gx_\mu\}, \{ \hat \gx_m\}$
in appendix \ref{gammaproperties}.  The eleven-dimensional spinors
decompose accordingly into a direct product of a four-dimensional
spinor and a seven-dimensional one. As we will assume that the
internal manifold $K_7$ has $G_2$ structure, we will mostly be
interested in the cases in which the internal spinor is the globally
defined spinor on such manifolds $\eta$ (which we also choose to be a
Majorana spinor)
\begin{equation}
  \label{redAnssp}
  \hat \Psi_M =
  (\hat \psi_M + \hat \psi_M^*) \otimes \eta \; .
\end{equation}
$\hat \psi_M$ is taken to be a Weyl spinor in order to agree with the
conventions which are mostly used in $N=1$ supergravity \cite{WB} and
hence we had to make the first term of the right hand side explicitly
a Majorana spinor. The field $\hat \psi_{\mu}$ in the above
decomposition will be the four-dimensional gravitino, while $\hat
\psi_m$ will be a spin $1/2$ field.

Finally we note that in order to have the $N=1$ supergravity properly
normalised one has to further perform field redefinitions such as the
following:
\begin{eqnarray}
  \label{resc}
  \hat g_{\mu \nu}  & = & \V^{-1} g_{\mu \nu} \; ,\nn \\
  \hat \gx_\mu & = & \V^{-1/2} \gx_\mu \; ,\\
  \hat \psi_\mu & = & \V^{-1/4} \psi_\mu \nn \; .
\end{eqnarray}
Note that the distinction between hatted and unhatted variables on the
internal manifold is not particularly important, and we will generally
omit hats where possible.

\subsection{$G_2$-structures}

For the following discussion it will be useful to point out a couple
of aspects of manifolds with $G_2$ structure. A more systematic
approach can be found in appendix \ref{g2structures}.

As we mentioned before, a manifold with $G_2$ structure admits a
globally defined spinor, $\eta$. This spinor fails in general to be
covariantly constant with respect to the Levi--Civita connection, but
one can always find a connection with torsion such that
\begin{equation}
  \label{cdeta1}
  \nabla_m^{(T)} \eta := \nabla_m \eta - \frac14 \kappa_{mnp} \gx^{np}
  \eta = 0 \; .
\end{equation}
The tensor $\kappa_{mnp}$ is the contorsion and gives a
characterisation of the $G_2$ structure as explained in appendix
\ref{g2structures}. Equivalently the $G_2$ structure can be described
in terms of the invariant three-form $\px$
\begin{equation}
  \label{phiDefn}
  \px_{mnp} := i \overline{\eta} \gx_{mnp} \eta  \; ,
\end{equation}
and its Hodge dual
\begin{equation}
  \label{PhiDefn} 
  \Phi_{mnpq} := (* \px)_{mnpq} = - \overline{\eta} \gx_{mnpq} 
  \eta \; . 
\end{equation}
Throughout this paper, we will use the properties of these forms as
set out in appendix \ref{g2structures}. For what follows it will be
useful to note that due to \eqref{cdeta1} the forms $\px$ and $\Phi$
are also covariantly constant with respect to the same connection with
torsion. Mathematically this means
\begin{equation}
  \label{cdphi1}
  \begin{aligned}
    \nabla_m^{(T)} \px_{npq} & := \nabla_m \px_{npq} - 3
    \kappa_{m[n}{}^r \px_{pq]r} = 0 \; , \\
    \nabla_m^{(T)} \Phi_{npqr} & := \nabla_m \Phi_{npqr} + 4
    \kappa_{m[n}{}^s \Phi_{pqr]s} = 0 \; . \\
  \end{aligned}
\end{equation}
This formula is quite important as it allows one to evaluate the
covariant derivative of these forms with respect to the Levi--Civita
connection in terms of the (con)torsion.

Among the different $G_2$ structures of a special role in our analysis
will be the so called manifolds with weak $G_2$ holonomy.  They are
characterised by the following relations (see also appendix
\ref{wg2hol} for more details)
\begin{equation}
  \label{wG2r}
  \begin{aligned}
    & d \px = \tau * \px \\
    & d * \px = 0 \; .
  \end{aligned}
\end{equation}
The quantity $\tau$ is a constant on the manifold with $G_2$ structure
which characterises completely the torsion of this type of manifolds.
In the context of M-theory compactifications, as we shall see later in
this paper, the parameter $\tau$ is not completely constant, but it
can depend on the space-time coordinates through the volume of the
internal manifold.

\section{Superpotential in M-theory compactifications on manifolds
  with $G_2$ structure}

\label{superpot}

Having described the general setup we can go and study in more detail
M-theory compactifications on manifolds with $G_2$ structure. In this
section we shall perform a general analysis which is valid for any
manifold with $G_2$ structure, and obtain information about the
K\"ahler potential and superpotential which appears in such
compactifications.

\subsection{Gravitino mass term}

In compactifying string and M-theories down to four dimensions it is
usually easier to derive the bosonic terms in the lower-dimensional
action. Computing fermionic terms is often more involved and the
fermionic part of the action is in general inferred from
supersymmetry.  There is, however, a specific class of terms which are
relatively easy to compute and which give valuable information about
the low energy effective action. These terms are the gravitino mass
terms, and in $N=1$ supergravity in four dimensions they take the form
\begin{equation}
  \label{m32}
  M_{3/2} = 
  \frac12 e^{K/2} \Big( \overline{W} \psi^T_\mu \gx^{\mu \nu} \psi_\nu + W
  \overline{\psi}_\mu \gx^{\mu\nu} \psi^*_\nu \Big) \; .
\end{equation}
$K$ here denotes the K\"ahler potential, which gives the coupling of
the chiral fields to supergravity, and $W$ is the superpotential. So
by computing such gravitino mass terms one can obtain information
about the K\"ahler potential and the superpotential. Similar
calculations were performed in \cite{BBHL,BBDG,GLM}.\footnote{The same
  information can be obtained by reducing the supersymmetry
  transformations as was done for example in \cite{BCtin}.}

Before we start the actual calculation, two comments about the
gravitino are in order. Firstly, $\psi_\mu$ is taken to be a Weyl
spinor to be consistent with \eqref{redAnssp}, and represents the
general result that any Majorana spinor $\chi$ can be expressed in
terms of a Weyl spinor $\psi$ by writing $ \chi = \psi + \psi^c$
(although in our conventions, charge conjugation ${}^c$ is equivalent
to complex conjugation ${}^*$).  Taking $\psi_\mu$ as positive
chirality gives $\psi_\mu^*$ as negative chirality.

Secondly, as we explained before, the relation \eqref{redAnssp} yields
both the four dimensional gravitino $\psi_\mu$ and also spin-$1/2$
fields $\psi_m$. Clearly, from the kinetic term in eleven dimensions
one obtains a kinetic term for the gravitino, one for the spin $1/2$
fields and one mixed kinetic term between the gravitino and the spin
$1/2$ fields. In the standard four-dimensional supergravity, such
mixed terms are not present and to obtain the correctly normalised
fermionic fields one has to perform a further redefinition of the
gravitino field which in terms of the eleven dimensional field
$\Psi_M$ takes the form
\begin{equation}
  \label{corrpsi}
  \Psi_\mu \to \Psi'_\mu = \Psi_\mu + \dx \Psi_\mu \; , \quad 
  \dx \Psi_\mu \sim \Gamma_\mu \Gamma^m \Psi_m \; .
\end{equation}
We note that, since $\psi_\mu$ in the above relation appears linearly,
the gravitino mass term \eqref{m32} for $\psi'_\mu$---when written in
terms of the uncorrected gravitino field $\psi_\mu$---has the same
form up to terms which are linear in the field $\psi_\mu$. Thus,
computing the terms $\psi_\mu \gx^{\mu \nu} \psi_\nu$, one can still
deduce the combination $e^{K/2} W$ and so in order to make the
calculation clearer we shall not be concerned with using the correct
definition for the gravitino field \eqref{corrpsi}.

We can now start to analyse the terms that contribute to the gravitino
mass term in four dimensions. We shall have in mind the most general
background compatible with Lorentz invariance in four dimensions,
which includes internal fluxes $F_{mnpq}$ and a flux in the four
space-time dimensions $F_{\mu \nu \rho \sigma}$. Note that the
background value for the fermionic fields is taken to vanish and so
the four-fermionic terms in the eleven-dimensional action cannot
contribute to the gravitino mass term in four dimensions.  Therefore,
we only need to consider the terms that we have kept in the action
\eqref{S11}, which are bilinear in the gravitino field.

In the following we shall analyse one by one these contributions to
the gravitino mass term.

\subsubsection{Contribution from the kinetic term}

As also remarked in \cite{GLM} the kinetic term for the gravitino only
produces a contribution to the mass term in the presence of
non-trivial structures, which will be proportional to $D_m \eta$.
Inserting the decomposition \eqref{redAnssp} in the gravitino kinetic
term from \eqref{S11} and keeping only the terms which are relevant
for the gravitino mass term one finds
\begin{eqnarray}
  \overline{\hat \Psi}_M \hat \Gx^{MNP} D_N \hat\Psi_P & = & 
  \overline{\hat \Psi}_\mu \hat \Gx^{\mu n \nu} D_n \hat\Psi_\nu 
  + \mathrm{\ terms\ not\ contributing\ to\ gravitino\ mass}
  \nn \\
  & = & - (\overline{ {\hat \psi}}_\mu + \overline{\hat \psi^*}_\mu) 
  \hat \gx^{\mu \nu} \gx
  (\hat \psi_\nu + \hat \psi_\nu^*) \eta^T \gx^n D_n \eta  
  + \ldots
\end{eqnarray}
We compute the covariant derivative on the spinor $\eta$ by
making use of \eqref{cdeta1} and we find that
\begin{equation}
  \label{rawKinetic}
  \eta^T \gx^n D_n \eta = - \frac{i}{4} \px^{mnp} \kx_{mnp} \; ,
\end{equation}
where $\kx_{mnp}$ denotes the intrinsic contorsion and the $G_2$
three-form $\px$ is defined in terms of the spinor $\eta$ in
\eqref{phiDefn}.  Note that this quantity picks up the singlet piece
under $G_2$ of the intrinsic contorsion and can be written using the
formulae in appendix \ref{g2structures} as
\begin{equation}
  \px^{mnp} \kx_{mnp} * \mathbf{1} = \frac{1}{2} d \px \wg \px \; .
\end{equation}
Taking into account the rescalings in \eqref{resc}, the contribution to
the gravitino mass coming from the kinetic term of the gravitino
in eleven dimensions can be written
\begin{equation}
  \label{M1}
  M_{3/2}^{(\mathrm{k.t.})} = \left(\frac{i}{16 \V^{3/2}} 
  \int d \px \wedge \px \right) \;
  \overline{\psi}_\mu \gx^{\mu \nu} \psi^*_\nu  + \mathrm{c.c.}
\end{equation}

\subsubsection{Contribution from the internal flux}

The next contribution to the gravitino mass term we discuss is the one
that appears due to the internal fluxes. Note that if one takes into
account a non-trivial $G_2$ structure the internal flux also receives
contribution from the non-closure of the forms in which we expand
the eleven-dimensional fields (see for example \cite{GLM}). We will
elaborate on this issue later and for now we denote the internal
fluxes by $\hat F_4$ and do not discuss their origin here. The
relevant term can be written
\begin{eqnarray}
  \overline{\hat \Psi}_M \hat \Gx^{MNPQRS} \hat \Psi_N (\hat F_4)_{PQRS} 
  & = & \overline{\hat \Psi}_\mu \Gx^{\mu \nu pqrs} \hat \Psi_\nu
  (\hat F_4)_{pqrs} + \ldots \nn \\
  & = & (\overline{ {\hat \psi}}_\mu + \overline{\hat \psi^*}_\mu)
  \hat \gx^{\mu \nu} (\hat \psi_\nu + \hat \psi_\nu^*)
  \eta^T \gx^{pqrs} \eta (\hat F_4)_{pqrs} + \ldots
\end{eqnarray}
Using \eqref{PhiDefn} to eliminate the $\eta^T \gx^{pqrs} \eta$,
taking into account all the factors in the action and the rescalings
from \eqref{resc}, the final result is
\begin{equation}
  \label{M2}
  M_{3/2}^{\mathrm{(i.f.)}} =\left(\frac{1}{8 \V^{3/2}} 
  \int \hat F_4 \wedge \px \right) 
  \;  \overline{\psi}_\mu \gx^{\mu \nu} \psi_\nu^* + \mathrm{c.c.}
\end{equation}

\subsubsection{Flux along the space-time directions}

Having a flux for $\hat F_4$ completely in the four space-time
directions can also generate a mass term for the gravitino from the
last term in the second line of \eqref{S11}.  After all rescalings are
taken into account, this would be simply given by
\begin{equation}
  \label{AC}
  M_{3/2}^{\mathrm{(e.f.)}} = - \frac12 \V^{3/2} d A_3 \wedge * C_4 \; ,
\end{equation}
where $C_4$ is defined as
\begin{equation}
  (C_4)_{\mu \nu \rho \sigma} = 3 (\overline{\psi} +
  \overline{\psi^*})_{[\mu} \gx_{\nu \rho} (\psi + \psi^*)_{\sigma]} =
  \frac{i}4 \left(\bar \psi_{\lx} \gx^{\lx \tau} \psi^*_\tau \right )
  \e_{\mu \nu \rho \sigma} + \mathrm{c.c.}
\end{equation}
As we will see in a short while, however, a purely four-dimensional
flux for $\hat F_4$ is not essential for obtaining some contribution
for the gravitino mass term from the above expression. The reason is
that we would have to dualise the three-form $A_3$ in a consistent
way.  A three-form in four dimensions is not dynamical and its dual is
thus only an arbitrary constant \cite{BPGG}. It is important to stress
that even if one chooses this constant to vanish the dualisation of
$A_3$ produces in general a non-trivial result due to its couplings in
the four-dimensional action. Thus, in order to obtain the correct
result we would have to derive first the complete action for the
four-dimensional field $A_3$. This requires that we make a specific
Ansatz for the decomposition of the eleven-dimensional field $\hat
A_3$ in terms of four-dimensional fields. At this stage we know almost
nothing about the correct way of performing such an expansion and so
consider
\begin{equation}
  \label{A3exp}
  \hat A_3 = \bgd{A}_3 + A_3 + a_3 \; ,
\end{equation}
where $\bgd{A}_3$ is the background value for $\hat A_3$, which gives
rise to the background flux, $G := d \bgd{A}_3$, while $A_3$ and $a_3$
represent fluctuations around it. $A_3$ is the four-dimensional
three-form field and $a_3$ is a three-form which lives in the internal
manifold, but which depends on the space-time as well. From the point
of view of the low energy action this form, $a_3$, will produce scalar
fields in four dimensions. Note that in general one can consider also
fluctuations which from the four-dimensional perspective are two-forms
or vector fields, but as these other fields play no role in the
following discussion we ignore them completely. However, it is
important to notice that apart from these other degrees of freedom,
\eqref{A3exp} is the most general expression one can consider. Once a
specific model is chosen the above Ansatz can be further refined (see
e.g. section \ref{deform}).

Inserting \eqref{A3exp} into \eqref{S11} and also considering the
rescalings \eqref{resc} one can derive the terms in the low energy
action which contain $A_3$. One is the kinetic term, and it is easy to
see that this has the form
\begin{equation}
  - \frac{\V^3}{4} d A_3 \wedge * d A_3 \; .
\end{equation}
The second contribution comes from the Chern-Simons term and is
present only if the internal value for the field $\hat F_4$ is
non-zero. As noted before this receives in general contributions from
two sources: from fluxes and from non-trivial structures. To make the
calculation clearer we will consider each case in turn.

\paragraph{Case I: fluxes on a manifold with $G_2$ holonomy} \ \\

On a manifold with $G_2$ holonomy the massless scalar fields which
come from $\hat A_3$ appear from the expansion in harmonic
three-forms.  Translated into our notation it means that we have to
take $a_3$ to be harmonic on the internal manifold. Turning on fluxes
in this case would mean that $\hat F_4$ has the following expansion
\begin{equation}
  \hat F_4 = G + d_4 A_3 + d_4 a_3 \; ,
\end{equation}
where the index $4$ on the exterior derivatives shows that they are
taken in the four space-time directions. $G$ is the four-form
background flux as defined above. It is not hard to see now that the
above expansion leads to the following expression for the
Chern--Simons coupling
\begin{equation}
  \label{CS-flux}
  \int_{K_7} \hat F_4 \wg \hat F_4 \wedge \hat A_3 = 6 \left( \int_{K_7}
  G \wedge a_3 \right ) \; d_4 A_3 \; .
\end{equation}

\paragraph{Case II: non-trivial $G_2$ structure and no fluxes}\ \\

As mentioned before, even if no fluxes are turned on one can in
general expect purely internal values for the field strength 
$\hat F_4$ from terms like $d_7 a_3$. With this in mind, the expansion
of the field strength $\hat F_4$ takes the form
\begin{equation}
  \hat F_4 = d_4 A_3 + d_4 a_3 + d_7 a_3 \; .
\end{equation}
The computation of the Chern--Simons term can now be seen to yield
\begin{equation}
    \label{CS-str}
  \int_{K_7} \hat F_4 \wedge \hat F_4 \wedge \hat A_3 = 3 \left(
  \int_{K_7} d_7 a_3 \wedge a_3 \right) \; d_4 A_3  \; ,
\end{equation}

As the above contributions to the action for $A_3$---\eqref{CS-str}
and \eqref{CS-flux}---are linear in $G$ and $d_7 a_3$, it is clear
that in the case where one turns on non-trivial fluxes in a
compactification on some manifold with $G_2$ structure the total
contribution to the term proportional to $d A_3$ in the
four-dimensional action is just the sum of the above
contributions. Taking into account the original term, \eqref{AC}, the
complete action for the field $A_3$ in four dimensions reads
\begin{equation}
  - \frac{\V^3}{4} d A_3 \wedge * d A_3 - \frac{\V^{3/2}}{2} d A_3
  \wedge * C_4 - \frac{1}{2} d A_3 \; \int_{K_7} \left( 
  G + \frac12 d_7 a_3 \right) \wg a_3 \; .
\end{equation}
Denoting the constant which is dual to $A_3$ by $\lambda$ the dual
action takes the form \cite{BPGG} (see also appendix E.2 of \cite{JLAM})
\begin{equation}
  \label{A3dual}
  - \frac{1}{4 \V^3} \left[ \lx - \int_{K_7}\left( G +
  \frac12 d_7 a_3 \right) \wg a_3 \right]^2 *\mathbf{1} + \frac12
  \left[ \lx - \int_{K_7} \left( G + \frac12 d_7 a_3
  \right) \wg a_3 \right]\V^{3/2} C_4 \; ,
\end{equation}
where we have neglected again four-fermionic terms.
The last term in the above expression is now truly the term which
contributes to the gravitino mass term.

\subsection{The superpotential}
\label{KW}

We are now in position to compute the superpotential which appears in
M-theory compactifications on seven-dimensional manifolds with $G_2$
structure. Putting together the contributions to the gravitino mass
term from \eqref{M1}, \eqref{M2} and \eqref{A3dual}, noting that the
flux $\hat F_4$ in \eqref{M2} is in fact $G + d_7 a_3$, one finds
\begin{equation}
  M_{3/2} = - \frac{i}{16 \V^{3/2}} \bar \psi_\mu \gx^{\mu \nu}
  \psi_\nu^* \left[\int (d a_3 + i d \px) \wedge (a_3 +  i \px) + 2\lx
  + 2 \int G \wedge (a_3 + i \px) \right] + \mathrm{c.c.}
\end{equation}
Using \eqref{m32} one immediately obtains (up to an overall phase
which plays no role in the definition of the superpotential)
\begin{equation}
  e^{K/2} W = \frac1{8 \V^{3/2}} \left[\int (d a_3 + i d \px) \wedge
  (a_3 +  i \px) + 2\lx + 2 \int G \wedge (a_3 + i \px) \right] +
  \mathrm{c.c.}
\end{equation}
This is the first main result of our paper so let us pause to discuss
it for a while. First of all we note that in $N=1$ supergravity, the
K\"ahler potential and the superpotential are not truly independent
functions, but the only thing which has a physical meaning is the
combination $e^K |W|^2$. Obviously the equation above gives us this
quantity. However we can not resist noting that this formula looks
very suggestive and that it splits up in a natural way in a part which
is holomorphic (the part in the brackets) and an overall real
factor. It is thus tempting to argue that for general
compactifications on manifolds with $G_2$ structure the K\"ahler
potential for the low energy fields is always given by the same
formula as in the case of $G_2$ holonomy
\begin{equation}
  \label{KG2}
  K = - 3 \ln \V \; .
\end{equation}
Establishing this, the superpotential is then given by
\begin{equation}
  \label{WG2}
  W = \frac18 \left[\int (d\px + i da_3) \wedge (\px +  i a_3) + 2\lx
  + 2 \int G \wedge (\px + ia_3) \right] \; .  
\end{equation}
One immediately observes in this superpotential that the last term is
precisely the result obtained in \cite{BW} for the case of manifolds
with $G_2$ holonomy. The second term, which is just a constant, should
always be present and appears from the correct dualisation of the
field $A_3$ in four dimensions. It also appears in \cite{BW}, but as a
quantum effect and in that case it turned out to be quantised. Here we
only limit ourselves to the supergravity approximation and thus we
keep this additional parameter real. Finally one notices the first
term which appears entirely due to the non-trivial $G_2$ structure
(intrinsic torsion). This term is completely new, and in the next
section we will compute it explicitly for the case of manifolds with
weak $G_2$ holonomy and also show that it reproduces the potential
which can be derived from the dimensional reduction when inserted in
the usual $N=1$ formula.

Before we end this section we make one more comment on the way this
superpotential was derived. In the original action for M-theory
\eqref{S11}, the fermion bilinears coupled only linearly to the Field
strength $\hat F_4$. Naively one would conclude from this that the
superpotential can depend only linearly on the flux $G$ or the fields
which appear in the low energy spectrum from $\hat A_3$. However it is
clear that the superpotential \eqref{WG2} contains also quadratic
terms like $G \wg a_3$ or $d a_3 \wg a_3$. Tracing back these terms we
see that they only appear from the correct dualisation of the three
form $A_3$ in four dimensions \eqref{A3dual}. The observation we want
to make is that there is a simpler and more direct way to see such
terms appearing by considering the starting action to be the `duality
symmetric' M-theory action \cite{BBS}. In such a formulation of
M-theory one also has a six-form field $\hat A_6$ with seven-form
field strength which is defined like
\begin{equation}
  \hat F_7 = d \hat A_4 + A_3 \wg F_4 \; .
\end{equation}
The fermionic action will now contain fermion bilinears which also
couple to $\hat F_7$ and from such couplings and also the second
term in the definition of $\hat F_7$ one can immediately see quadratic
terms in $\hat A_3$ appearing in the superpotential.

\section{Metric deformation space of manifolds with weak $G_2$ holonomy}

\label{deform}

So far we have only discussed general features of M-theory
compactifications on manifolds with $G_2$-structure without making any
reference to the low-energy field content of such theories. In order
to be able to obtain a specific model in four dimensions one needs to
have some more information about the internal properties of such
manifolds and in particular one needs to get a handle on their moduli
space.%
\footnote{Strictly speaking the word modulus/moduli stand for fields
which are exactly flat directions of the potential. What we actually
mean by moduli in this section and also in the whole paper are the
fields which appear from the metric deformations on the internal
manifold. Generically, as we shall see in a while, such fields appear
with a potential and it is precisely this feature we are interested to
capture here, namely determine the potential for such fields which
would enable one to perform a correct analysis of the field
stabilisation in such models.} %
In general this question is quite complicated and a satisfactory
answer has not been found yet. For special cases, however, such as the
one we will discuss here, it turns out to be possible to gain
information about the space of deformations of these manifolds and
using this information to compute the low-energy action in four
dimensions. It will be our purpose in this section to solve some of
the problems outlined here for the case of manifolds with weak $G_2$
holonomy.

\subsection{Deformations of the weak $G_2$ structure}

In this section and also in the following ones we will use many
properties of manifolds with $G_2$ structures. For the reader who is
not familiar with these notions we have assembled a couple of
appendices where we present the relations we will use in these
sections. We will only outline the key facts about the deformation
space of manifolds with weak $G_2$ holonomy and prove many of the
underlying (rather important) facts in the appendix \ref{wg2hol}.

We start by noting that specifying a $G_2$ structure $\px$ on a
seven-dimensional manifold uniquely determines the metric.  The
relevant equation for this is \eqref{gphi}, but for our purposes it
suffices to note that the variations of the metric can be encoded in
variations of the $G_2$ three-form $\px$.  Thus, as in the case of
$G_2$ holonomy, the deformations of metrics on manifolds with $G_2$
structure can be studied by looking at variations of the invariant
form $\px$. For a general $G_2$ structure the situation is a bit
complicated because the torsion itself can vary together with the
structure. For this reason, from now on we shall concentrate on the
special class of manifolds with weak $G_2$ holonomy \eqref{wG2r} (for
a detailed description see appendix \ref{wg2hol}).  From \eqref{wG2r}
it is clear that the torsion $\tau$ cannot depend explicitly on the
coordinates of the internal manifold, but from what was said above, it
can in principle depend on its moduli and as we shall see later, in
our case it does.

The strategy we adopt is the following. We consider first a $G_2$
structure $\px$ which satisfies the weak $G_2$ conditions
\eqref{wG2r}. Then we consider small variations of this $G_2$
structure by some arbitrary form $\dx \px$ and impose that the
equations \eqref{wG2r} are satisfied for some $\tau' = \tau + \dx
\tau$.  This will yield some conditions on the variations $\dx \px$
and $\dx \tau$.  It is important to stress here that by changing
$\px$, the metric on the manifolds changes as well and thus the
definition of the Hodge star changes. In order not to create any
confusion we will use the notations from \cite{Joyce}. Proposition
10.3.5 from \cite{Joyce} gives the form of the Hodge dual of a
perturbed $G_2$ structure to be
\begin{equation}
  \label{Theta}
  \Theta(\px + \dx \px) = * \px + \frac43 * P_1 \dx \px + * P_7 \dx \px
  - * P_{27} \dx \px \; ,
\end{equation}
where $\Theta(\xi)$ is a map $\Theta : \Lambda^3 \to \Lambda^4$ which
computes the Hodge dual of a three-form $\xi$ with the metric defined
by $\xi$ via \eqref{gphi}.  The Hodge star on the right hand side of
\eqref{Theta} is defined from the old (unvaried) metric are $P_1$,
$P_7$ and $P_{27}$, which denote projection operators on the spaces of
corresponding dimensionality.  Note that on the space of three-forms
$P_1 +P_7 + P_{27} = \unitop$.

Imposing the first relation in \eqref{wG2r} for the varied form $\px +
\dx \px$ at the first order in the perturbations we can write
\begin{equation}
  \label{wG2pr}
  d \dx \px = \dx \tau *\px + \tau \left( \frac43 * P_1 \dx \px + * P_7
  \dx \px - * P_{27} \dx \px \right)  \; .
\end{equation}
Thus, perturbing a weak $G_2$ structure $\px$ by some form $\dx \px$
leads again to a weak $G_2$ structure provided the variation $\dx \px$
satisfies \eqref{wG2pr} for some suitable $\dx \tau$.

At this stage, we make use of the observation that the $\rep{7}$
component of $\dx \px$ makes no contribution to perturbations of the
induced metric through the formula \eqref{deltag}.  We shall therefore
set such components to zero, which will simplify the analysis below.
It is shown in the Appendix that for forms which satisfy $P_7 \dx \px
=0$, the projectors $P_1$ and $P_{27}$ commute with the exterior
derivative. Since these projectors also commute with the Hodge star
one can break \eqref{wG2pr} into two simpler conditions for the
singlet variations $P_1 \dx \px$ and for the ones which transform as a
$\rep{27}$ under $G_2$
\begin{equation}
  \label{wG2split}
  \begin{aligned}
    d P_1 \dx \px & = \dx \tau * \px + \frac43 \tau * P_1 \dx \px \; , \\
    d P_{27} \dx \px & = - \tau * P_{27} \dx \px \; .
  \end{aligned}
\end{equation}
From here it is clear that the torsion $\tau$ depends on the
deformations of the weak $G_2$ manifold only via the singlet
deformation $P_1 \dx \px$. In other words, since such singlet
deformations only rescale the $G_2$ structure $\px$ and through it the
volume of the manifold, we conclude that the torsion $\tau$ depends on
the parameters describing the weak $G_2$ manifold only via its volume.

Let us now consider the two equations above separately. We start with
the first one and parameterise the singlet part of the deformation as
\begin{equation}
  \label{singdef}
  P_1 \dx \px = \e \px \; ,
\end{equation}
for some arbitrary small $\e$. Then equation \eqref{wG2split} becomes
\begin{equation}
  \e d \px = \dx \tau * \px + \frac43 \tau \e *\px \; .
\end{equation}
Using again the weak $G_2$ condition \eqref{wG2} we obtain
\begin{equation}
  \dx \tau = - \frac{\epsilon}{3} \tau \; .
\end{equation}
From the definition of the volume in terms of the $G_2$ structure
$\px$ \eqref{vol} and \eqref{Theta}, it is not hard to see that under
the deformation \eqref{singdef} the volume changes as
\begin{equation}
  \dx \V = \frac73 \e \V \; .
\end{equation}
Dividing the last two equations we obtain that the variation of the
torsion $\tau$ with the volume obeys
\begin{equation}
  \label{dtdv}
  \frac{\dx \tau}{\tau} = - \frac17 \; \frac{\dx \V}{\V} \; ,
\end{equation}
or after integration
\begin{equation}
  \tau \sim \V^{-1/7} \; .
\end{equation}
Intuitively the above equation can be understood as follows. As
mentioned before singlet variations rescale the $G_2$ structure
$\px$. In its turn, such a rescaling produces a rescaling of the
metric via \eqref{deltag} and so the scalar curvature $R$ is
rescaled. Since the torsion is directly related to the scalar
curvature via \eqref{Rmn} it follows that such a deformation can only
be present if the torsion of the manifold itself changes and the
quantitative measure of this change is captured by the above equation.

Let us now turn our attention to the second equation in
\eqref{wG2split}. As we said before, in the case of variations with
forms which transform as $\rep{27}$ under $G_2$ the torsion does not
vary. Intuitively, since the torsion is a singlet under $G_2$ we would
need another object which transforms non-trivially under $G_2$ in
order to produce a singlet out of the deformation $\dx \px$. Since we
do not have at our disposal any such thing it can be understood that
the torsion can not change under such deformations. Thus, the second
equation in \eqref{wG2split} only imposes a condition on the
variations of $\px$ which are compatible with the weak $G_2$
structure.

So far we have learned that the non-trivial metric deformations of
weak $G_2$ manifolds are parameterised by three-forms $\dx \px$
satisfying
\begin{equation}
  \label{dpxwg2}
  \begin{aligned}
    P_7 \dx \px & = 0 \; ,\\
    d P_{27} \dx \px & = - * P_{27} \dx \px \; .
  \end{aligned}
\end{equation}
Note the counter-intuitive minus sign appearing on the right hand side
which is going to be crucial in determining the correct mass of the
modes associated with these variations of the $G_2$ structure.

Let us now try to give a more explicit parameterisation of the
deformation space. Recall that for the case of manifolds with $G_2$
holonomy the form $\px$ was closed and coclosed and thus harmonic.
Consequently one expanded this form in a basis for harmonic
three-forms with the coefficients being the moduli fields.  Then by
general methods, which we present in appendix \ref{g2structures}, one
could compute the metric on the moduli space. Let us try to do
something similar here. Clearly from the relations \eqref{wG2r} we see
that the form $\px$ cannot be harmonic anymore and in fact the torsion
$\tau$ measures its failure to be harmonic. However, it is easy to
compute the action of the Laplace operator on $\px$ and one obtains
\begin{equation}
  \label{Lapphi}
  \Delta \px \equiv (*d*d + d*d*) \px = \tau^2 \px \; ,
\end{equation}
and thus $\px$ is still an eigen-form of the Laplace operator
corresponding to the eigenvalue $\tau^2$. Consider a basis $\Pi_i$ for
the three-forms satisfying
\begin{equation}
  \Delta \Pi_i\equiv (*d*d + d*d*) \Pi_i = \tau^2 \Pi_i \; ,
\end{equation}
where these forms do not depend on the moduli of the weak $G_2$
manifold. As it stands this assumption is definitely too strong.  In
fact what happens for harmonic forms is that their dependence on the
moduli comes in only via exact forms. Assuming the same thing happens
also in our case, we show in the appendix that the calculation which
follows from here on is indeed consistent. For now we can expand the
$G_2$ structure $\px$ in this basis
\begin{equation}
  \label{pexp}
  \px = s^i \Pi_i \, ,
\end{equation}
and as for manifolds of $G_2$ holonomy we think of the forms $\Pi_i$
as independent of the choice of metric. For the parameters $s^i$ to be
truly moduli of the weak $G_2$ structure, we still need that the forms
$\Pi_i$ satisfy the condition \eqref{dpxwg2}.  From here on we will
assume that the forms $\Pi_i$ used in the expansion \eqref{pexp} do
satisfy this condition as well. If this is true, than the coefficients
$s^i$ are indeed the scalar fields which characterise the possible
deformations of the weak $G_2$ manifold.

As one does on a manifold with $G_2$ holonomy, let us further define
\begin{equation}
  \label{ki}
  \cK_i = \int \Pi_i \wg * \px \; .
\end{equation}
Using the fact that the forms $\Pi_i$ satisfy \eqref{dpxwg2} one
derives 
\begin{eqnarray}
  d \Pi_i & = & d (P_1 \Pi_i + P_{27} \Pi_i)  \nn \\
  & = & d P_1 \Pi_i - \tau * P_{27} \Pi_i \\
  & = & - \tau * \Pi_i + 2 d P_1 \Pi_i \nn \; .
\end{eqnarray}
The projector on the singlet subspace $P_1$ is defined by
\begin{equation}
  P_1 \Pi_i = \frac{1}{7 \V} \int \Pi_i \wg * \px = \frac{\cK_i}{7 \V}
  \; ,
\end{equation}
where we have used that $(\Pi_i)_{mnp} \px^{mnp}$ does not depend on
the internal manifold, fact which is also proven in the appendix.
Using the weak $G_2$ conditions \eqref{wG2r} one immediately finds
\begin{equation}
  \label{dpi}
  d \Pi_i = - \tau * \Pi_i + \frac{2 \tau \cK_i}{7 \V} * \px \; .
\end{equation}

This is the central relation of this second part of the paper as it
will allow us to compute explicitly the low-energy effective action
for compactifications on manifolds with weak $G_2$ holonomy and
compare it with the result derived before on general grounds.

It is important to stress one more thing here. At this stage we cannot
say much more about the space of deformations of weak $G_2$ manifolds,
apart from the fact that it can be characterised in terms of the forms
$\Pi_i$ satisfying \eqref{dpxwg2}, which will be exploited in the next
section. One thing is, however, quite important.  Note that the
operator $\Delta - \tau^2$ is an elliptic operator. It is known in the
general theory of operators that elliptic operators on compact
manifolds have a finite dimensional kernel. This means that at least
the space of deformations of weak $G_2$ manifolds is
\emph{finite}.\footnote{We thank Andr\'e Lukas for drawing our
  attention on this fact.}

Before we move on let us summarise the main results that we derived in
this section. For weak $G_2$ manifolds the invariant form $\px$ turns
out to be an eigenform of the Laplace operator \eqref{Lapphi}.
Performing the expansion in terms of the deformations in the same way
as one did on manifolds with $G_2$ holonomy we conclude that the forms
in which we perform this expansion must obey \eqref{dpxwg2}. It is
interesting to note that this relation was obtained in a pretty
generic fashion. Similar relations were derived for half-flat
manifolds \cite{GLMW} but using mirror symmetry as an additional
source of information for the manifolds with $SU(3)$ structure.
Moreover those relations were valid only in a certain limit which was
denoted as the small torsion limit while the relations above are valid
for any torsion. We will see later on that the torsion cannot be small
as it turns out to be of the order of the inverse radius of the
manifold.

\subsection{Useful formulae on the deformation space of weak $G_2$
  manifolds}

\label{formwg2}

Before we perform the compactification we will find it useful to
derive some formulae which make the calculation on the deformation
space of weak $G_2$ manifolds easier.

The presence of the forms $\Pi_i$ which are not closed (although they
are co-closed) allows us to introduce a topological, two-index,
symmetric object on these manifolds
\begin{equation}
  \label{Kdef}
  \cK_{ij} = \int \Pi_i \wg d \Pi_j = \cK_{ji} \; .
\end{equation}
Obviously, the appearance of such a matrix is only due to the
non-minimal structure as it depends on $d \Pi_i$, which would clearly
vanish for the case of manifolds with $G_2$ holonomy. As we shall see
later on this object will enter the expression of the superpotential
in terms of the low energy fields.

A straightforward calculation, which we have outlined in the appendix,
shows that for a general expansion of the form \eqref{pexp} the sigma
model metric for the moduli takes the form
\begin{equation}
  \label{gdef}
  g_{ij} = \frac{1}{4 \V} \int \Pi_i \wg * \Pi_j \; .
\end{equation}
Using  \eqref{dpi} and \eqref{Kdef} it is easy to show that
\begin{equation}
  g_{ij} = - \frac{1}{4 \tau \V} \cK_{ij} + \frac{\cK_i \cK_j}{14 \V^2} \; .
\end{equation}
Furthermore one also has the usual relations
\begin{eqnarray}
  \cK_i s^i & = & 7 \V \; , \nn \\
  \cK_i & = & 4 \V g_{ij} s^j \; , \\
  \cK_i g^{ij} & = & 4 \V s^j \; . \nn
\end{eqnarray}

The matrix $\cK_{ij}$  introduced in \eqref{Kdef} can be shown to
satisfy 
\begin{equation}
  \begin{aligned}
    \cK_{ij} s^j & = \tau \cK_i \; ,\\
    \cK_{ij} g^{jk} & = - 4 \tau \V \dx_i^k + \frac87 \tau \cK_i s^k
    \; .
  \end{aligned}
\end{equation}

Using these relations we can now proceed and compute the effective
action which arises by compactifying M-theory on manifolds with weak
$G_2$ holonomy.

\section{M-theory compactifications on manifolds with weak $G_2$
  holonomy}

Having discussed in the previous section the possible deformations of
weak $G_2$ manifolds we shall now move on and derive the low energy
action which appears when compactifying M-theory on such manifolds.
Then we will show that the resulting theory is an $N=1$ supergravity
coupled to chiral multiplets. The correspnding K\"ahler potential and 
superpotential will turn out to be the ones derived on general grounds
in section \ref{KW}.

\subsection{Review of the Freund--Rubin solution}

In order to understand properly the process of the compactification it
is useful to review the Freund--Rubin reduction Ansatz which turns out
to be the relevant setup for compactifications on manifolds with weak
$G_2$ holonomy. Recall that in the Freund--Rubin background, only the
flux along the four space-time directions is non-vanishing and
constant and the space-time is considered to be maximally symmetric.
The unique possibility for the background is then
\begin{equation}
  \label{FRbg}
  F_{\mu \nu \rho \sigma} = \frac32 \tau \e_{\mu \nu \rho \sigma} \; ,
\end{equation}
where the factor $3/2$ is chosen for later convenience.  We know that
such a background is supersymmetric if the supersymmetric variation of
the gravitino vanishes. Inserting \eqref{FRbg} into \eqref{susytrs}
and splitting the eleven-dimensional spinor like in section
\ref{superpot}, one immediately finds that for the internal manifold
\begin{equation}
  \label{spFR}
  \nabla_m \eta + \frac{i}{8} \tau \gx_m \eta = 0 \; . 
\end{equation}
This is precisely the weak $G_2$ condition in terms of the spinor
$\eta$ \eqref{deleta} which was shown in the appendix \ref{wGi} to be
equivalent to \eqref{wG2r}.  We note that the most general
seven-dimensional manifold that produces a maximally symmetric
space-time and preserves supersymmetry in the background \eqref{FRbg}
is a weak $G_2$ manifold\footnote{For a rigorous proof of this
statement see \cite{BJ}.}  satisfying \eqref{spFR} or equivalently
\eqref{wG2r}.  Consequently looking for small deformations of this
solution that still satisfy the equations of motion in the background
\eqref{FRbg} is equivalent to looking for small variations of the
metric on the weak $G_2$ manifold that lead to another weak $G_2$
manifold with some torsion $\tau + \dx \tau$.  One can see that the
parameters $s^i$ obtained from the deformations of the weak $G_2$
structure will indeed have the interpretation of scalar fields in four
dimensions.

In general Kaluza--Klein compactifications one first identifies the
massless modes and truncates away the massive towers of modes which
appear. For this it is necessary to identify correctly the mass terms
for the various fields in the theory.

For compactifications on $G_2$ manifolds, which are Ricci flat, this
is a straightforward exercise and the masses of the different modes
can be obtained by studying the spectrum of the Laplace operators
acting on various degree forms and of the Lichnerowicz operator. For
the case at hand the situation is a bit more complicated because of
the fact that manifolds with weak $G_2$ holonomy have a non-vanishing
Ricci curvature. This implies that in general the ground state of such
a theory is no longer Minkowski, but AdS. In particular the Ricci
tensors for the external and internal spaces are respectively
\begin{equation}
  \label{Ric}
  R_{\mu \nu} = - \frac{3}{4} \tau^2 g_{\mu \nu} \; , \hspace{10mm}
  R_{mn} = \frac38 \tau^2 g_{mn} \; .
\end{equation}
Note that the setup we are dealing with is significantly changed from
what was considered before in the literature. The torsion in our case
is finite and in particular related to the radius of the internal
manifold. It is not arbitrary small as, for example, was previously
assumed for half-flat manifolds \cite{GLMW}. Thus one has to take into
account all the large effects coming from torsion and fluxes in order
to determine the correct ground state. The general analysis for the
compactification of M-theory on seven-dimensional Einstein manifolds
was performed in \cite{DNP}, and so we will just adapt the formul\ae\
they use for our own purposes.

In our analysis we will only be interested in the scalar fields which
appear in the four-dimensional effective action. Such mass operators
were computed for example in \cite{DNP} and for the conventions we use
in this paper they take the form
\begin{equation}
  \label{mop}
  \begin{aligned}
    M_{0^-}^2 & = ~Q^2 + \frac32 \tau Q + \frac12 \tau^2
    = (Q+ \tau)(Q + \frac12 \tau) , \\
    M_{0^+}^2 & = ~\Delta_L - \frac14 \tau^2 \; ,
  \end{aligned}
\end{equation}
where we have defined as in \cite{DNP} the operator $Q := * d$.  The
former of these operators is for three-form matter fields and the
latter for traceless symmetric variations of the metric.  For
three-form structure variations that generate such metric variations
via \eqref{deltag}, this operator can---after a tedious calculation
involving the Lichnerowicz operator $\Delta_L$---be written
\begin{equation}
  M_{0^+}^2 = Q(Q + \frac12 \tau) \; .
\end{equation}
The presence of supersymmetry should, however, complexify the $G_2$
structure $\px$ by the matter three-form $a_3$ and therefore to
preserve $N=1$ supersymmetry in four dimensions one needs to expand
$a_3$ in the same way as $\px$.  It is not hard to see using the above
formulae for the masses of the scalars coming from the metric
deformations and $A_3$ matches the mass pattern of a Wess-Zumino
(chiral) multiplet in AdS space~\cite{BG}, confirming our expectations
about supersymmetry. Finally we note that the forms satisfying
\eqref{dpxwg2} turn out to be zero-modes of the operator
$M_{0^-}^2$. 

\subsection{The compactification}

To perform the compactification on such manifolds with weak $G_2$
holonomy one has first to identify the fields which appear in four
dimensions. In the previous section we have argued that the AdS
massless modes (scalars) which appear in compactifications on
manifolds with weak $G_2$ holonomy are given by the expansion in forms
which satisfy \eqref{dpxwg2}. Neglecting as in the usual Kaluza--Klein
setups the rest of the massive towers of states we can now perform the
compactification on weak $G_2$ manifolds and keep only the modes
discussed above.

From the expansion \eqref{pexp} and the relations \eqref{gphi} one can
derive what will be the kinetic term for the scalars $s^i$ which comes
from the expansion of the Ricci scalar. As in the case of manifolds
with $G_2$ holonomy the sigma-model metric takes the form
\begin{equation}
  \label{gmods}
  g_{ij} = \frac{1}{4 \V} \int \Pi_i \wg * \Pi_j \; .
\end{equation}

In the matter sector we perform a similar expansion to \eqref{pexp}.
In this paper we will only be interested in the scalar fields which
arise in the compactification of M-theory on a manifold with weak
$G_2$ holonomy. There can be also other fields like vectors, but here
we will ignore them completely.  If we denote again as in
\eqref{A3exp} the internal component of the field $A_3$ by $a_3$, then
we write
\begin{equation}
  \label{a3expand}
  a_3 = a^i \Pi_i \; .
\end{equation}
The full eleven-dimensional three form $\hat A_3$ then takes the form
\begin{equation}
  \label{A3expand}
  \hat A_3 = A_3 + a^i \Pi_i \; ,
\end{equation}
where $A_3$ is a three-form in four dimensions. This is not dynamical
and so it can be dualised to a constant as we did in section
\ref{superpot}.  The four-dimensional bosonic action which one derives
in this way has the form
\begin{equation}
  \label{S4}
  S_4 = \frac12 \int\left[ \sqrt{-g} R -  g_{ij} d T^i \wg * d \bar
    T^j - \sqrt {-g} V  \right] \; .
\end{equation}
The potential $V$ comes from three distinct places. First it comes
from the purely internal part of of $\hat F_4$. This will have the
form
\begin{equation}
  V_1 = \frac{1}{8 \V^2} \int d a_3 \wg * d a_3 \; ,
\end{equation}
where the exterior derivative is understood to be in the internal
manifold direction and the factor $1/\V^2$ comes from the Weyl
rescaling in four dimensions, \eqref{resc}. Using \eqref{dpi} this can
be easily seen to be
\begin{equation}
  \label{V1}
  V_1 = \frac{\tau^2}{8 \V^2} a^i a^j \int \Pi_i \wg * \Pi_j  =
  \frac{\tau^2}{2 \V} a^i a^j g_{ij} \; .
\end{equation}
From the dualisation of $A_3$ in four dimensions we have already seen
that there is a contribution to the potential \eqref{A3dual}. Using
\eqref{a3expand} and \eqref{dpi} one finds%
\footnote{Note that we now take the dual of $A_3$ in four dimensions
  to be the constant $\lx$, which can in principle be independent of
  the background value considered in \eqref{FRbg}.}%
\begin{equation}
  \label{V2}
  V_2 = \frac{1}{4 \V^3} \left( \lambda - \frac{a^i a^j \cK_{ij}}{2}
  \right)^2 \; .
\end{equation}
Finally one has to take into account the contribution from the
curvature of the internal manifold. The Ricci scalar of weak $G_2$
manifolds can be easily computed \eqref{Rmn}, and after performing the
integration over the internal manifold whilst taking into account the
factor $1/\V^2$ coming from the Weyl rescaling in four dimensions one
obtains
\begin{equation}
  \label{V3}
  V_3 = - \frac{21 \tau^2}{16 \V} \; .
\end{equation}

The potential coming from the compactification thus takes the form
\begin{eqnarray}
  \label{Vtot}
  V & = & V_1 + V_2 + V_3 \nn \\
  & = & \frac{1}{16} \left[ -21 \frac{\tau^2}{\V} + \frac{1}{\V^3}
    (a^i a^j \cK_{ij})^2 + 16 \frac{\tau^2}{\V} a^i a^j g_{ij} \right]
  \; .
\end{eqnarray}

\subsection{Comparison with the general result}

To conclude this analysis we still have to show that the result
obtained in the previous subsections is indeed an $N=1$
supergravity. As we have neglected completely the gauge fields we only
have to find the corresponding K\"ahler potential and
superpotential. This is not a hard task since we have at our disposal
the general result derived in section \ref{KW}. In appendix
\ref{g2structures} it was shown that the metric \eqref{gmods} is
K\"ahler, i.e.  $g_{ij} = \partial_i \partial_{\bar \jmath} K$, and
the K\"ahler potential is
\begin{equation}
  \label{Kpot}
  K = -3 \ln \V \; .
\end{equation}
As we argued in section \ref{KW}, the superpotential is given by
\eqref{WG2}, which for the case where a non-trivial structure but no
$G$ fluxes are taken into account becomes
\begin{equation}
  W = \frac18 \int_{M_7} d (a_3 + i \px) \wg (a_3 + i \px) +
  \frac{\lambda}{2} \; .
\end{equation}
Using the field expansions \eqref{pexp} and \eqref{a3expand} and also
the definition \eqref{Kdef} we obtain the superpotential in terms of
the four-dimensional fields
\begin{equation}
  \label{WwG2}
  W = \frac{\cK_{ij}}{8} T^i T^j + \frac{\lambda}{2} \; ,
\end{equation}
where the complex fields $T^i$ are defined as
\begin{equation}
  \label{Ti}
  T^k = a^k + i s^k \; .
\end{equation}

To show that the action \eqref{S4} is the bosonic part of an $N=1$
supergravity theory we have to show that the potential \eqref{Vtot}
can be derived from the superpotential \eqref{WwG2} using the general
supergravity formula
\begin{equation}
  V = e^K \left[ D_i W \overline{(D_j W)} 
    g^{\bar \jmath i} - 3 |W|^2 \right] \; ,
\end{equation}
where as usual $D$ denotes the K\"ahler covariant derivative.

The calculation is a bit tedious, but completely straightforward and
so we will present only the main steps in the following. First one can
derive
\begin{equation}
  D_i W = \frac14 \cK_{ij} T^j + \frac{i}{2} \; \frac{\cK_i}{\V} W \; .
\end{equation}
Using now the formulae from section \ref{formwg2} one easily finds that
\begin{equation}
  D_i W \overline{(D_j W)} g^{\bar \jmath i} = \frac14 (- \tau \V \cK_{ij} +
  \frac27 \tau^2 \cK_i \cK_j ) T^i \bar T^j - \frac{i}{2} \tau \cK_i
  (T^i \bar W - \bar T^i W) + 7 |W|^2 \; .
\end{equation}
Furthermore, one shows that
\begin{equation}
   \frac{i}{2} \tau \cK_i (T^i \bar W - \bar T^i W) = -7 \tau \V Re \;
   W + 4 (Im \; W)^2 \; .
\end{equation}
Finally, one obtains
\begin{equation}
  4 |W|^2 - \frac{i}{2} \tau \cK_i ( T^i \bar W - \bar T^i W) =
  \frac{1}{16} \left[ (a^i a^j \cK_{ij})^2 - (s^i s^j \cK_{ij})^2 \right] \; .
\end{equation}

Putting all the results together one obtains the final form of
the potential
\begin{equation}
  V = \frac{1}{16} \left[ -21 \frac{\tau^2}{\V} + \frac{1}{\V^3}
    (a^i a^j \cK_{ij})^2 + 16 \frac{\tau^2}{\V} a^i a^j g_{ij} \right]
  \; ,
\end{equation}
which is precisely the potential derived in \eqref{Vtot} from the
compactification side.

To conclude, we have shown in this section that the compactification
of M-theory on a manifolds with weak $G_2$ holonomy leads to an $N=1$
supergravity coupled to chiral superfields in four dimensions with
K\"ahler potential defined by \eqref{Kpot} and superpotential
\eqref{WwG2}. This is a nice test of the general analysis of M-theory
compactifications on manifolds with $G_2$ structure presented in
section \ref{superpot} where the superpotential was derived from
computing the four-dimensional gravitino mass term.

\subsection{Inclusion of non-vanishing flux}

In the previous section, we didn't consider the contributions to the
superpotential arising from the internal flux $G$, since this term has
already been covered for $G_2$ holonomy in \cite{WB} and we do not
expect this analysis to be any different in our case. It is still,
however, necessary to consider the forms in which it will be
appropriate to expand the flux $G$. Firstly, note that following the
discussion of the previous section, we have expanded the three-form
leading to four-dimensional scalars as
\begin{equation}
  T_3 = \px + i a_3 = T^i \Pi_i \; .
\end{equation}
It is then clear that the part of the superpotential arising from
$G$-flux in \eqref{WG2} takes the form
\begin{equation}
  W_{\mathrm{flux}} = \frac{1}{4} \int G \wg T_3
  = \frac{1}{4} \langle *G , T \rangle \; ,  
\end{equation}
where $\langle , \rangle$ denotes the inner product for forms.
Clearly, this quantity will vanish unless $*G$ has the same eigenvalue
of $Q = *d$ as $T$ does.%
\footnote{Note that the operator $Q$ acting on 3-forms on a
  seven-dimensional manifold is self-adjoint as $Q^\dagger =
  (d*)^\dagger = * d^\dagger = **d* = d* = Q$.}%
This suggests na\"{\i}vely that we should expand $G$ in the $* \Pi_i$,
however these forms will not in general be independent of the choice
of the metric. It should nevertheless be possible to find linear
combinations of the $* \Pi_i$ that form a dual basis $\{
\widetilde{\Pi}^i \}$ obeying
\begin{equation}
  \int \Pi_i \wg \widetilde{\Pi}^j = \dx_i^j \; .  
\end{equation}
We then expand $G = 4 G_i \widetilde{\Pi}^i$.  Note that turning on
this flux will in general either break supersymmetry or introduce
warping into the compactification, so it is usual to consider $G$ as
somehow `small'.

This analysis produces a final form for the superpotential of
\begin{equation}
  W = \Lambda + G_i T^i + k_{ij} T^i T^j \; ,
\end{equation}
for constant $\Lambda, G_i, k_{ij}$. The study of superpotentials of
this form arising from compactification of M-theory on manifolds of
$G_2$ structure is of phenomenological interest in its own right and
is a question that we hope to return to in a later publication.  In
particular we would like to extend the analysis of \cite{BAS} to see
if vacua of small or vanishing cosmological constant can be obtained
for the case of weak $G_2$ compactifications of M-theory.

\section{Conclusions}

In this paper we have analysed M-theory compactifications on manifolds
with $G_2$ structure. Using the globally defined spinor which exists
on such manifolds one can define the four-dimensional gravitino and
compute explicitly the terms which give rise to the gravitino mass
term in four dimensions. From this we were able to derive the general
form for the superpotential which appears in M-theory
compactifications on manifolds with $G_2$ structure \eqref{WG2}. This
formula generalises in a natural way the one which was derived for
manifolds with $G_2$ holonomy in \cite{BW} which was derived based on
the conjecture made in \cite{GVW}.

However, even if such a compact formula can be written for the
superpotential, its expression in terms of the low energy fields is
not known unless one specifies further the structure of the internal
manifold. This was the purpose of the second part of this paper where
we derived the effective action that appears from compactifications of
M-theory on manifolds with weak $G_2$ holonomy. It turns out that the
possible metric variations on such manifolds are in one to one
correspondence with the three-forms on the weak $G_2$ manifold
satisfying
\begin{equation}
  d \ax = - \tau * \ax \; .
\end{equation}
It also turns out that fluctuations of the three-form field $\hat A_3$
that are proportional to such forms lead to scalar fields in four
dimensions that are massless in the background AdS solution. Thus, as
for the case of manifolds with $G_2$ holonomy, the metric fluctuations
compatible with the structure and the massless modes of the three-form
field $\hat A_3$ pair up into complex fields which will be the scalar
components of the chiral superfields in four dimensions. The
superpotential appears to be quadratic in the superfields \eqref{WwG2}
and we have shown by an explicit calculation that the potential which
can be derived from this superpotential matches the one which appears
from the compactification on the weak $G_2$ manifold.

The phenomenological viability of the model we have constructed here
is not very clear at the moment, mainly because of the large AdS
curvature of the four-dimensional space. However, there are several
directions which are worth investigating.  The first of these would be
a systematic study of the mass operators \eqref{mop} in AdS
backgrounds, which may also admit stable states of mass$^2 \leq
0$ that we have not considered here.  Secondly the appearance of a
quadratic superpotential opens up a new set of possibilities for
finding a minimum of the potential as well as BPS states in the
four-dimensional supergravity such as \cite{HL}.

Furthermore, explicit examples of weak $G_2$ compactifications may be
relevant in the process of obtaining the Standard Model spectrum from
M-theory. It is well known that for chiral fermions to appear one
needs conical singularities on the internal space
\cite{AW,AcW}. However, until now there was no explicit construction
of a \emph{compact} manifold with $G_2$ holonomy which contains such
singularities. On the other hand, the only explicit example of a
compact manifold with $G_2$ structure which has such singularities is
the case constructed in \cite{BM} and which is a weak $G_2$ manifold.
Considering an explicit example would be interesting both in terms of
looking for supersymmetric minima and BPS states in that model and
also for studying anomaly cancellation.

In view of the work in \cite{BC1,BC2} it would be interesting to
perform a similar analysis to that above in the case of weak $SU(3)$
holonomy, where we expect to find analogous results.

\vspace{1cm}
\noindent
{\large\textbf{Acknowledgments:}} This work was supported by PPARC.
It is a pleasure to thank Klaus Behrndt, Andrew Bruce, Andr\'e Lukas,
Paul Saffin and Daniel Waldram for useful conversations. We also thank
Michael Haack for pointing us an important reference at an early stage
of this work, and Annamaria Kiss for a careful reading of the manuscript.

\vspace{1cm}
\noindent
{\LARGE\textbf{Appendix}}

\appendix

\section{Conventions and notation}

\label{gammaproperties}

Throughout the paper we use the following notation. Uppercase indices
$M,N, \ldots = 0, \ldots , 10$ denote curved eleven-dimensional
indices. Four-dimensional indices are denoted by Greek letters $\mu,
\nu , \ldots = 0 , \ldots , 3$ and we use lowercase letters $m,n,
\ldots = 1 , \ldots , 7$ for the indices on the internal
manifold. Finally, indices $i,j \ldots$ are used to label the
directions on the deformation space of manifolds with weak $G_2$
holonomy. 

Where possible we use the index-free (form) notation. The relation the
tensor notation is given by
\begin{equation}
  A_{(p)} = \frac{1}{p!} A_{m_1 \ldots m_p} dx^{m_1} \wg \ldots
  dx^{m_p} \; .
\end{equation}
In our conventions, the standard operations on forms
in $D$ dimensions are given by
\begin{eqnarray}
  (d A)_{m_1 \ldots m_{p+1}} & = & 
  (p+1) \nabla_{[m_1} A_{m_2 \ldots m_{p+1}]} \nn \\
  (d^{\dagger} A)_{m_1 \ldots m_{p-1}} & = & 
  - \nabla^{n} A_{n m_1 \ldots m_{p-1}}  \\
  (*A)_{m_1 \ldots m_{D-p}} & = & \frac{(-1)^{p(D-p)}}{p!}
  \epsilon_{m_1 \ldots m_{D-p}}{}^{n_1 \ldots n_p} A_{n_1 \ldots n_p} \nn
\end{eqnarray}
where we use the conventions that the $\epsilon$-symbol is a proper
tensor (rather than a tensor density) which is normalised as
\begin{equation}
  \e_{0 1 \ldots  D-1} = + \sqrt { \left| \DET (g) \right| } \; .
\end{equation}
The symbol $g$ usually denotes the metric as follows: $g_{MN}, \
g_{mn}$ and $g_{\mu \nu}$ are space-time metrics while $g_{ij}$
denotes the metric on the deformation space.

Finally we use the following conventions for the gamma-matrices.
The eleven-dimensional gamma-matrices $\{ \Gx_M \}$ satisfy the
Clifford algebra
\begin{equation}
  \big\{ \Gx_M , \Gx_N \big\} = 2 g_{MN} \; ,
\end{equation}
and are taken to be real. Furthermore, they are chosen to satisfy
\begin{equation}
  \Gx_0 \Gx_1 \ldots \Gx_{10} = 1 \; ,
\end{equation}
or equivalently
\begin{equation}
  \e^{M_1 \ldots M_{11}} \Gx_{M_1} \ldots \Gx_{M_{11}} = - 11! 
\end{equation}
According to \eqref{split} we decompose the gamma matrices as
\begin{eqnarray}
{\Gx}_{\mu} & = & {\gx}_{\mu} \otimes \unitop \; ,\nn \\
{\Gx}_m & = & \gx \otimes {\gx}_m \; .
\end{eqnarray}
$\{ \gx_\mu \} $ are the four-dimensional
gamma matrices which are also chosen to be real, while $\{\gx_m\}$
are the gamma matrices on the internal manifold and
are chosen purely imaginary. Note that this choice of reality of the
gamma matrices is consistent with the fact that the the
eleven-dimensional ones are real as the four-dimensional chirality
matrix $\gx$ is defined as
\begin{equation}
  \gx = i {\gx}_0 \ldots {\gx}_3 \in i \mathbb{R} \; ,
\end{equation}
and is purely imaginary. Last, the gamma matrices $\gx_m$ on the
internal manifold satisfy
\begin{equation}
  \gx_1 \ldots \gx_7 = i \; .
\end{equation}
Our conventions for spinor conjugation are that, for general spinor
$\psi$, in Minkowskian-signature spaces we have 
\begin{equation}
  \overline{\psi} :=  \psi^{\dagger} \gamma^0 \; ,   
\end{equation}
where $^{\dagger}$ denotes Hermitian
conjugation and in Euclidean-signature spaces 
\begin{equation}
  \overline{\psi} :=\psi^{\dagger} \; .   
\end{equation}

\section{$G_2$ structures}

\label{g2structures}

In this appendix we review the most important features about $G_2$
structures which we need in the calculations in the paper.

\subsection{General properties}

\label{g2general}

We consider the manifolds with $G_2$ structure to be seven dimensional
manifolds which admit a globally defined, nowhere-vanishing spinor
which we denote by $\eta$. 
Without restricting the generality of the setup we consider
this spinor to be Majorana and we normalise it as
\begin{equation}
  {\overline \eta} \eta = 1 \; .
\end{equation}
It is well-known that one can always find a connection---which in
general will not be torsion-free---that makes the globally defined spinor
covariantly constant:
\begin{equation}
  \label{cdeta}
  D_m^{(T)} \eta = D_m \eta - \frac14 \kappa_{mnp} \gx^{np}
  \eta = 0 \; ,
\end{equation}
where $D_m$ denotes the covariant derivative with respect to the
Levi-Civita connection and $\kappa_{mnp}$ is the contorsion. 

Using the spinor $\eta$ one can construct a globally defined and
nowhere vanishing totally antisymmetric tensor
\begin{equation}
  \px_{mnp} = i \eta^T \gx_{mnp} \eta \; ,
\end{equation}
where $\gx_{mnp}$ denotes the antisymmetric product of three gamma
matrices with unit norm. The Hodge dual of the form $\px$ can be
written similarly
\begin{equation}
  \Phi_{mnpq} = (* \px)_{mnpq} = \frac{i}{3!} \epsilon_{mnpq}{}^{rst}
  \eta^T \gx_{rst} \eta = - \eta^T \gx_{mnpq} \eta \; .
\end{equation}
It is straightforward to check using \eqref{cdeta} that both $\px$ and
$\Phi$ are covariantly constant with respect to the connection with
torsion, i.e.
\begin{equation}
  \label{cdphi}
  \begin{aligned}
    \nabla_m^{(T)} \px_{npq} & := \nabla_m \px_{npq} - 3
    \kappa_{m[n}{}^r \px_{pq]r} = 0 \; , \\
    \nabla_m^{(T)} \Phi_{npqr} & := \nabla_m \Phi_{npqr} + 4
    \kappa_{m[n}{}^s \Phi_{pqr]s} = 0 \; , \\
  \end{aligned}
\end{equation}
Note that these are the only forms which can be constructed from
spinor bilinears as the combinations with antisymmetric products of
one, two, five and six gamma matrices vanish identically. Using the
$\px$ one can write the volume of the manifold with $G_2$ structure
to be
\begin{equation}
  \label{vol}
  \V = \frac17 \int_{K_7} \px \wg * \px = \frac17 \int_{K_7} \px
  \wg \Phi \; .
\end{equation}

In addition to the covariantly constant spinor, $\eta$, there are
seven spinor directions left which we denote $\eta_m$. These
obey
\begin{eqnarray}
\overline \eta \eta_m = {\overline \eta_m} \eta & = & 0 \; , \nn \\
{\overline \eta_m} \eta^n & = & \delta_m^n \; . \label{etaortho}
\end{eqnarray}
The action of antisymmetrised products of gamma matrices on $\eta$
is given by
\begin{eqnarray}
\gamma^{m} \eta & = & i g^{mn} \eta_n \nn \\
\gamma^{mn} \eta & = & \px^{mnp} \eta_p \nn \\
\gamma^{mnp} \eta & = & -i \px^{mnp} \eta + i \Phi^{mnpq} \eta_q \nn \\
\gamma^{mnpq} \eta & = & - \Phi^{mnpq} \eta
- 4 \px^{[mnp } g^{ q ] r} \eta_r \\
\gamma^{mnpqr} \eta & = & -i \left( \Phi^{[ mnpq} g^{r]s} 
+ 4 \px^{[mnp} \px^{qr] s} \right) \eta_s \nn \\
\gamma^{mnpqrs} \eta & = & - \left( \Phi^{[mnpq} \px^{rs]t}
+ 4 \px^{[mnp} \Phi^{qrs]t} \right) \eta_t \nn \\
\gamma^{mnpqrst} \eta & = & i \e^{mnpqrst} \eta
- i \left( \Phi^{[mnpq} \Phi^{rst]u}
+4 \px^{[mnp} \Phi^{qrs}{}_v \px^{t]uv} \right) \eta_u \; . \nn
\end{eqnarray}
By using the above, together with \eqref{etaortho},
Fierz identities and Dirac matrix identities such as those
in \cite{Candelas:1984yd} one can derive some useful formulae relating
the forms $\px$ and $\Phi$
\begin{eqnarray}
  \label{G2id}
  \px_{mnr} \px^{rpq} & = & \Phi_{mn}{}^{pq} + 2 \delta_{mn}^{pq} \\
    \px^{mns} \Phi_{spqr} & = & 
    6 \delta^{\left[m\right.}_{\left[p\right.} 
    \px^{\left.n\right]}{}_{\left.qr\right]} \label{phiPhi} \\
  \Phi_{mnpt} \Phi^{qrst} & = & 6 \delta^{mnp}_{qrs}
  + 9 \Phi_{\left[mn\right.}{}^{\left[qr\right.}  
  \delta^{\left.s\right]}_{\left.p\right]}
  - \px_{mnp} \px^{qrs} \label{mnpqrs} \\
  36 \delta^{\left[ rs\right.}_{\left[ mn \right.}
  \px_{\left. pq\right]}{}^{\left. t \right]} & = &
  \Phi_{mnpq} \px^{rst} + 4 \px_{\left[ mnp \right.} 
    \Phi_{\left. q \right]}{}^{rst} - \e_{mnpq}{}^{rst} \\
  24 \px^{[tu}{}_{(m} \delta^{v]}_{n)} & = &
  \e^{pqrstuv} \px_{mpq} \px_{nrs}  \\
  \e_{mnpqrst} & = & 5 \px_{[mnp} \Phi_{qrst]} \label{ephiPhi} \; ,
\end{eqnarray}
where $\delta_{m_1 \ldots m_a}^{n_1 \ldots n_a} : = 
g_{\left[ m_1 \right.}{}^{\left[ n_1 \right.} \cdots 
g_{\left. m_a \right]}{}^{\left. n_a \right]}$. 
Further identities may be derived by contracting indices in the above.

\subsection{Classification of $G_2$ structures}

As we said before, manifolds with $G_2$ structure are characterised by
the existence of a globally defined spinor or equivalently by a three
form $\px$. Also, as we shall see in the next section, the structure
defines a metric via \eqref{gphi}.  However, in general, neither the
spinor $\eta$ nor the three-form $\px$ are covariantly constant with
respect to the Levi--Civita connection defined by this metric. The
failure of this connection to preserve the structure defines the
intrinsic torsion and measures the deviation of the $G_2$ structure
from the $G_2$ holonomy case. It is well known \cite{Joyce} that in
general the intrinsic torsion is a one-form with values in the
orthogonal complement of the structure group in $SO(n)$. For the
structure group $G_2$ this can be encoded in the following expression
\begin{equation}
  T^0 \in \W_1 \oplus \W_2 \oplus \W_3 \oplus \W_4 \; ,
\end{equation}
where the description of the torsion classes $\W_1 , \ldots , \W_4$ is
given in the table \ref{Tcls}. We have used the superscript 0 to
denote degrees of freedom `left-over' from other projections.
\begin{table}[h]
  \centering
  \begin{tabular}[h]{|c|c|c|}
    \hline
    \quad Class \quad \raisebox{-3mm}{\rule{0pt}{8mm}} & \quad
    Dimension \quad & \quad Interpretation \quad \\ 
    \hline
    $\W_1$ \raisebox{-3mm}{\rule{0pt}{8mm}} & $\rep{1}$ & 
    $d \px \llcorner \Phi $ \\
    \hline
    $\W_2$ \raisebox{-3mm}{\rule{0pt}{8mm}} & $\rep{7}$ & 
    $\px \llcorner d \px, \px \llcorner d*\px $ \\ 
    \hline
    $\W_3$ \raisebox{-3mm}{\rule{0pt}{8mm}} & $\rep{14}$ & 
    $(d * \px)^0 $ \\
    \hline
    $\W_4$ \raisebox{-3mm}{\rule{0pt}{8mm}} & $\rep{27}$ & 
    $(d \px)^0$ \\
    \hline
  \end{tabular}
  \caption{Description of the torsion classes of manifolds with $G_2$
    structure.} 
  \label{Tcls}
\end{table}

Using the $G_2$ invariant form, $\px$,
and its Hodge dual $\Phi = * \px$, one can 
define projection operators which
select some particular representation of the group $G_2$. In
particular we will be mostly interested into the projector on the
singlet subspaces. For example the torsion component which is in
$\W_1$ is defined as
\begin{equation}
  (P_1 T)_{mnp} = T^1 \px_{mnp} \; ,
\end{equation}
where 
\begin{equation}
  T^1 = \frac{\int \sqrt \g \; T_{mnp} \px^{mnp}}{\int \sqrt \g \;
  \px_{mnp}\px^{mnp}} = \frac{1}{42 \V} \int \sqrt \g \; T_{mnp}
  \px^{mnp} \; .
\end{equation}
Using \eqref{cdphi} and \eqref{phiPhi} one can compute
\begin{eqnarray}
  d \px \wg \px = d \px \wg * \Phi & = & \frac{\sqrt{g}}6 \nabla_{[m}
  \px_{npq]} \Phi^{mnpq} = \frac{\sqrt{g}}2 \kappa_{[mn}{}^r
  \px_{pq]r} \Phi^{mnpq} \nn \\
  & = & 2 \sqrt{g} \kappa_{mnp} \px^{mnp} = 2 \sqrt{g} T_{mnp} \px^{mnp} \; ,
\end{eqnarray}
and so $T^1$ takes now the form
\begin{equation}
  T^1 = \frac{1}{84 \V} \int d \px \wg \px \; .
\end{equation}

\subsection{Induced metric}

As we pointed out before, a globally defined three-form on a
seven-dimensional space assures that the manifold has $G_2$
structure. Moreover, given such a three-form $\px$, a metric is also
uniquely defined by 
\begin{equation}
  \label{gphi}
  \begin{aligned}
    g_{mn} = & \left(\DET (s)\right)^{-1/9} s_{mn} \; , \\
    s_{mn} = & \frac{1}{144} \px_{mp_1 p_2} \px_{np_3 p_4} \px_{p_5 p_6
    p_7} \hat \e^{p_1 \ldots p_7} \; ,
  \end{aligned}
\end{equation}
where $\hat \e^{m_1 \ldots m_7} = \sqrt{\DET (g)} \e^{m_1 \ldots m_7}
= \pm 1$. Thus one can parameterise the deformations of the metric
$\dx g_{mn}$ in terms of the deformations of the $G_2$-structure 
$\dx \px$. Using the above definition of the metric and the relations
\eqref{phiPhi} one immediately finds
\begin{equation}
  \dx s_{mn} = \frac{\sqrt{\DET (g)}}{2} \px_{(m}{}^{pq} \dx \px_{n)pq} \; .
\end{equation}
With this the metric variation $\dx g_{mn}$ becomes
\begin{equation}
  \label{deltag}
   \dx g_{mn} = \frac{1}{2} \px_{(m}{}^{pq} \dx \px_{n)pq}
  - \frac{1}{18} ( \px \llcorner \dx \px ) g_{mn} \; . 
\end{equation}
Here, as elsewhere in this paper, $\llcorner$ denotes the contraction
of indices, so that for $p$-form $\ax$ and $q$-form $\bx$ obeying
$q<p$,
\begin{equation}
  ( \bx \llcorner \ax )_{m_1 \ldots m_{q-p}} :=
  \bx_{n_1 \ldots n_q} \ax^{n_1 \ldots n_q}{}_{m_1 \ldots m_{q-p}} \; .
\end{equation}

\subsection{Metric on the deformation space}

Let us suppose that we have expanded $\px$ in terms of some parameters $s^i$
\begin{equation}
  \label{pexp1}
  \px = s^i \Pi_i \; ,
\end{equation}
where $\Pi_i$ are three forms which further satisfy 
\begin{equation}
  \label{noP7}
  P_7 \Pi_i =0  
\end{equation}
and which are supposed not to depend on the parameters $s^i$. The
meaning of the above equation is that using the set of forms $\Pi_i$
one can parameterise the deformations of the $G_2$-structure $\px$ in
terms of variations of the parameters $s^i$. These parameters, or more
correctly their variations, are constant on the internal manifold, but
from a four-dimensional perspective they become (scalar) fields. We
have seen in the previous section that variations of the $G_2$
structure induce variations in the metric on the manifold via
\eqref{deltag} and so we can say that the parameters $s^i$ describe
the metric fluctuations on the internal manifold. Consequently the
kinetic term for these fields in four dimensions appears from the
expansion of the eleven-dimensional Ricci scalar. It is well-known
that this leads to the following term
\begin{equation}
  \int_{M_{11}} \sqrt{-g} R_{11} = \int_{M_{11}}
  \sqrt{-g} \V \left\{ R_4 + R_7 + \frac{1}{4\V} \left[ \partial_{\mu}
  g_{mn} \partial^{\mu} g^{mn} - \mathrm{tr}(\partial_{\mu} g)
  \mathrm{tr}(\partial^{\mu} g) \right] \right\} \; .
\end{equation}
Inserting \eqref{deltag} in the above equation and using
\eqref{mnpqrs} one obtains
\begin{eqnarray}
  \label{KaluzaKij}
  S_{\mathrm{kin}} & = & - \frac{1}{2} \int_{M_4} 
  \sqrt{-g} \partial_{\mu} s^i \partial^{\mu} s^j 
  \int_{K_7} \sqrt{g} \frac14 
  \left[ \left( \frac{1}{2} \px_{(m}{}^{pq} \dx \px_{n)pq}
      - \frac{1}{18} ( \px \llcorner \dx \px ) g_{mn} \right)^2
    -  \frac{1}{81} (\px \llcorner \Pi_i ) (\px \llcorner \Pi_j )
  \right] \nn \\
  & = & - \frac14 \int_{K_7} \Pi_i \wg * \Pi_j \int_{M_4} \sqrt{-g} \;
  \partial_{\mu} s^i \partial^{\mu} s^j 
  +\frac32 \int_{M4} \sqrt{-g} \; \frac{\partial_{\mu} \V \partial^{\mu}
    \V}{\V^2}  \; .
\end{eqnarray}
To read off the sigma-model metric for the scalars $s^i$ in four
dimensions one has to take into account the redefinitions of the
space-time metric by a volume factor \eqref{resc}. Note that under
this rescaling the last term in the above equation disappears and we
are left with
\begin{equation}
  \label{sigmet}
  g_{ij} = \frac{1}{4 \V} \int_{K_7} \Pi_i \wg * \Pi_j \; .
\end{equation}
Since the resulting four-dimensional action is supposed to be
supersymmetric, the above metric has to be K\"ahler. This is indeed
the case and the K\"ahler potential was derived on general grounds in
section \ref{KW}
\begin{equation}
  \label{Kpot1}
  K = - 3 \ln (\V) \; ,
\end{equation}
where the volume $\V$ was defined in \eqref{vol}. To show that this is
the K\"ahler potential corresponding to the metric \eqref{sigmet} we
have to know the dependence of the the volume on the parameters $s^i$.
$\px$ depends linearly on $s^i$, \eqref{pexp1} provided the forms
$\Pi_i$ are independent of these parameters. For the form $\Phi$ this
dependence is more complicated because of the Hodge duality operation
which is involved, but its variation with the parameters $s^i$ can
be read off from \eqref{Theta}. One finds
\begin{equation}
  \label{varV}
  \frac{\dx \V}{\dx s^i} = \frac17 \int_{K_7} \Pi_i \wg \Phi +
  \frac17\int_{K_7} \px \wg * \frac43 \Pi_i = \frac13 \int_{K_7} \Pi_i
  \wg \Phi \; .
\end{equation}
With this, one immediately finds the first derivative\footnote{Note
  that a K\"ahler potential makes sense only in the context of
  complex geometry. Thus what we have in mind here is that the
  K\"ahler potential \eqref{Kpot1} is a function of the complex fields
  $T^i$ defined in \eqref{Ti} and thus derivatives are then taken with
  respect to the fields $T^i$ rather then only their imaginary parts
  $s^i$.} of the K\"ahler potential \eqref{Kpot1} 
\begin{equation}
  \begin{aligned}
    K_i := \partial_i K & = \frac{1}{2} \frac{\partial}{\partial s^i}
    \left( - 3 \ln (\V) \right) \nn \\
    & =  \frac{-1}{2 \V} \int \Pi_i \wg \Phi \label{Ki} \; .
  \end{aligned}
\end{equation}
Using again the relation \eqref{Theta} we can compute the second
derivative of the K\"ahler potential
\begin{eqnarray}
  \label{Kij}
  K_{i \bar \jmath} & = & \frac{1}{4} \left( 
    \frac{\partial \V}{\partial s^j} \V^{-2} \int \Pi_i \wg \Phi
    - \V^{-1} \int \Pi_i \wg \left( \frac{4}{3} * P_1 \Pi_j
      - * P_{27} \Pi_j \right) \right) \nn \\
  & = & \frac{1}{4 \V} \int \Pi_i \wg * \Pi_j \; ,
\end{eqnarray}
where we have used \eqref{noP7} and the last equality used
\eqref{varV}. As anticipated one notices 
now that the metric \eqref{sigmet} can be indeed derived from the
K\"ahler potential \eqref{Kpot1}. We should stress here that this
result is quite general and holds as long as the forms $\Pi_i$ do not
depend on the parameters $s^i$ and they satisfy \eqref{noP7}.

\section{Weak $G_2$ holonomy}

\label{wg2hol}

Usually not much can be said about generic manifolds with $G_2$
structure. The characterisation of such manifolds needs in general the
introduction of other forms besides $\px$ and $\Phi$ in order to
parameterise the torsion (see for example \cite{GLMW}). However there
is a simple case where such manipulations are not necessary. This is
the case of manifolds with weak $G_2$ holonomy which are characterised
by the fact that the intrinsic torsion resides completely within the
first torsion class (see table \ref{Tcls}), or in other words, it is a
singlet under the $G_2$ structure group. We can thus write the
contorsion
\begin{equation}
  \label{ct}
  \kappa_{mnp} = \kappa \px_{mnp} \; ,
\end{equation}
where $\kappa$ is a constant on the manifold with $G_2$ structure.
Note that this is the only possibility we have as $\px_{mnp}$ is the
unique three-index object which is a singlet under $G_2$. It is clear
that the contorsion is totally antisymmetric and thus it also
coincides with the torsion tensor $T_{mnp} = \kappa_{[mn]p}$. Using
\eqref{ct} and the fact that $\px$ is covariantly constant with
respect to the connection with torsion we can compute the exterior
derivative of $\px$
\begin{equation}
  \label{contDeriv}
  (d \px)_{mnpq} = 4 \nabla_{[m} \px_{npq]} = 12 
  \kappa_{[mn}{}^r \px_{pq]r} = 12 \kappa \Phi_{mnpq} 
  \; .
\end{equation}
We thus see, as anticipated in the table \ref{Tcls}, that the exterior
derivative of $\px$ is indeed again a singlet, namely $\Phi$. It will
be more convenient to introduce another parameter $\tau = 12 \kappa $
such that
\begin{equation}
  \label{ctwG2}
  \kappa_{mnp} = \frac{\tau}{12} \px_{mnp} \; .
\end{equation}
in terms of which the weak $G_2$ condition takes the more custom form
\begin{equation}
  \label{wG2}
  \begin{aligned}
    & d \px =  \tau *\px \; , \\
    & d * \px = 0 \; .
  \end{aligned}
\end{equation}

Having defined the general properties of manifolds with $G_2$
structure we can now analyse the main features of the special case of
manifolds with weak $G_2$ holonomy in the following.

\subsection{Weak $G_2$ identities}
\label{wGi}

Starting from \eqref{cdphi} and using the relations \eqref{phiPhi} one
can show that
\begin{equation}
  \label{delpx}
  \nabla_m \px_{npq} = 3 \kappa_{m[n}{}^r \px_{pq]r} = \frac{\tau}{4}
  \Phi_{mnpq} \; .
\end{equation}
Repeating the procedure for $\Phi$ one finds
\begin{equation}
  \label{delPhi}
  \nabla_m \Phi_{npqr} = -4 \kappa_{m[n}{}^s \Phi_{pqr]s} = -
  \frac{\tau}{3} \phi_{m[n}{}^s \Phi_{pqr]s} = - \tau g_{m[n} 
  \px_{pqr]} \; .
\end{equation}
From the above relation one immediately derives 
\begin{equation}
  \label{ddpx}
  \Box \px_{mnp} 
  = \nabla_q \nabla^q \px_{mnp} 
  = - \frac{\tau^2}{4} \px_{mnp} \; .
\end{equation}

In the main text we also needed a couple of relations involving the
curvature tensors of weak $G_2$ manifolds. To compute these we start
from the fact that the globally defined spinor is covariantly constant 
with respect to the connection with torsion \eqref{cdeta}.
Since we work with imaginary gamma matrices the spinor $\gx^{pq} \eta$
is still a Majorana spinor and thus can be expanded in terms of a
basis for Majorana spinors on the weak $G_2$ manifold 
$\{\eta, \eta_m \}$ as defined in \S \ref{g2general}. 
It is then straightforward to derive that
\begin{equation}
  \label{deleta}
  D_m \eta = \frac{1}{4} \kappa_{mnp} \gx^{np} \eta =  
  \frac{\tau}{8} \eta_m \; .
\end{equation}
Taking the commutator of two covariant derivatives acting on the
spinor $\eta$ one obtains
\begin{equation}
  \label{Reta}
  R_{mnpq} \gx^{pq} \eta = \frac{\tau^2}{8} \gx_{mn} \eta \; .
\end{equation}
Multiplying this from the left with $\gx^n$ and using the first Bianchi
identity for the Riemann tensor immediately gives the Ricci curvature 
of weak $G_2$ manifolds as
\begin{equation}
  \label{Rmn}
  R_{mn} \equiv R^p{}_{mpn} = \frac38 \tau^2 g_{mn} \; ,
\end{equation}
which shows that these manifolds are Einstein. Note, as a matter of
fact, that the scaling behavior of $\tau$ found in \eqref{dtdv} is in
agreement with the above relation.

Multiplying again \eqref{Reta} by $\gx^s$ and contracting with
$\eta^T$ one obtains
\begin{equation}
  \label{Rpx} 
  R_{mnpq} \px^{pqs} = \frac{\tau^2}{8} \px_{mn}{}^s \; ,
\end{equation}
which can be used to prove other identities about the
Riemann tensor in weak $G_2$.

\subsection{More about weak $G_2$ manifolds}

There are some more properties of weak $G_2$ which are rather
important for the analysis presented in the paper and which we still
have to prove. We will mainly be interested in three-forms which seem
to play a key role in the manifolds we were discussing. Before we
start there is a useful remark we should make about three-forms on
seven-dimensional manifolds (see for example \cite{DNP}): The
eigenfunctions of the Laplace operator corresponding to a non-zero
eigenvalue $\mu^2$ are in one to one correspondence with the
eigenfunctions of the operator $Q = *d$ corresponding to the
eigenvalues $\pm \mu$. As the reader has probably already noticed the
operator $Q$ plays an important role in our analysis as the forms
which are relevant for the deformations of weak $G_2$ manifolds are
eigenfunctions of this operator corresponding to the eigenvalue $-
\tau$. It is these forms we are going to analyse in the following. As
mentioned in the main text, the forms we are interested in also do not
contain a part which transforms under $\rep{7}$ under $G_2$. It would
be interesting if one could prove the existence of such forms, but
this goes beyond the scope of this paper and so for our purposes we
will just assume that the forms with these desired properties do
indeed exist.

Let us consider the forms $\Pi_i$ as in the main text which satisfy
\begin{equation}
  \label{proppi}
  \begin{aligned}
    \Delta \Pi_i & = \tau^2 \Pi_i \; , \\
    P_7 \Pi_i & = 0 ~ \Leftrightarrow ~ (\Pi_i)_{mnp} \Phi^{mnp}{}_r = 0
    \; , 
  \end{aligned}
\end{equation}
and prove as we have stated that for such forms the projectors $P_1$
and $P_{27}$ commute with the differential operator $Q = *d$.
For this, we consider the quantity
\begin{equation}
  (\Pi_i)_{mnp}  \px^{mnp} \; ,
\end{equation}
and show that it does not depend on the coordinates of the internal manifold.
To see this we compute
\begin{equation}
  \nabla_m (\px_{npq} (\Pi_i)^{npq}) = \frac\tau4 \Phi_{mnpq}
  (\Pi_i)^{npq} + \px_{npq} (d \Pi^i)_m{}^{npq} + 3 \px_{npq} \nabla^n
  (\Pi_i)_m{}^{pq} \; .
\end{equation}
Using \eqref{proppi}, the first term clearly vanishes. Furthermore,
we can choose without loss of generality that the forms $\Pi_i$ are
eigenfunctions of the $Q= *d$ operator with eigenvalue $\pm \tau$.
With this the second term becomes proportional to 
$\Phi_{mnpq} (\Pi_i)^{npq}$ which again vanishes for the forms we
consider. Thus, we are left with
\begin{equation}
  \nabla_m (\px_{npq} (\Pi_i)^{npq}) = 3 \px_{npq} \nabla^n
  (\Pi_i)_m{}^{pq} \; .
\end{equation}
The right hand side can be computed if we push $\px$ through the
derivative as $d *\px = 0$ and then notice that the combination
$\px_{npq} (\Pi_i)_m{}^{pq}$ is symmetric in the indices $m , \ n$
which is again a consequence of \eqref{proppi}. We obtain
\begin{equation}
  \nabla_m (\px_{npq} (\Pi_i)^{npq}) = 3 \nabla^n(\phi_{npq}
  (\Pi_i)_m{}^{pq}) = 3 \nabla^n(\phi_{mpq} (\Pi_i)_n{}^{pq}) = 3
  \frac\tau4 \Phi^n{}_{mpq} (\Pi_i)_n{}^{pq} \; ,
\end{equation}
which again vanishes upon using \eqref{proppi}. Note that in the last
relation we have also used that $d * \Pi_i = 0$ which holds true if we
take the forms $\Pi$ to be eigenfunctions of the operator $Q$. This
completes the proof of
\begin{equation}
  \nabla_m (\px_{npq} (\Pi_i)^{npq}) = 0 \; .
\end{equation}
Since $P_1 \Pi_i$ is defined as
\begin{equation}
  (P_1 \Pi_i)_{mnp} = \frac1{42} (\px_{qrs} (\Pi_i)^{qrs}) \px_{mnp}
  \; ,
\end{equation}
we conclude that
\begin{equation}
  [P_1 , Q] \Pi_i = 0 \; .
\end{equation}
Since for these forms it also holds that $P_1 \Pi_i + P_{27} \Pi_i =
\Pi_i$ it follows that
\begin{equation}
  [P_{27} , Q] \Pi_i = 0 \; .
\end{equation}

There is one more aspect which is crucial in the whole construction
which we did in this paper, namely the dependence on the parameters
$s^i$ introduced in \eqref{pexp} of the basis of forms we consider
$\Pi_i$. In principle there is no reason to believe that the solutions
of the equation $\Delta = \tau^2$ are independent of the parameters
$s^i$ of the manifold as the metric itself depends on them. In fact
when one does a variation of the metric the operator $\Delta$ changes
and so we expect its solution to change as well. This also happens in
ordinary manifolds with restricted holonomy like Calabi--Yau manifolds
or manifolds with $G_2$ structure. In these cases however, one can
easily show that such a dependence on the moduli of the harmonic forms
is exact. If one assumes that the same happens for the case of forms
which are eigenvalues of the Laplace operator corresponding to some
non-zero eigenvalue then it is quite easy to show that such a `mild'
dependence on the parameters is not going to affect the results we
have derived so far. First of all it is straightforward to see that
this dependence drops out completely from the definition of
$\cK_{ij}$. The other thing to show is that the metric on the
deformation space does not get an additional dependence on the
parameters from the forms $\Pi_i$. Indeed, if such a dependence on the
parameters $s^i$ of these forms is only via an exact form one can
immediately see that
\begin{equation}
  \int \dx \Pi_i \wg * \Pi_j = \int d \bx_i \wg * \Pi_j = - \int \bx_i
  \wg d* \Pi_j = 0 \; ,
\end{equation}
because the forms $\Pi_i$ are coclosed. We thus conclude that the only
relevant dependence on the parameters of the weak $G_2$ manifold
$s^i$, is via the expansion \eqref{pexp} as we considered in the main text.

\end{document}